\documentclass[useAMS,usenatbib,aas_macros]{mn2e}
\usepackage{aas_macros}
\usepackage{amsmath}
\usepackage{graphicx}
\usepackage{color}
\newcommand{\adb}[1]{\textcolor{black}{#1}}

\newcommand{\equ}[1]{eq.~(\ref{eq:#1})}

\newcommand{\se}[1]{\S\ref{sec:#1}}
\newcommand{\fig}[1]{Fig.~\ref{fig:#1}}

\newcommand{\Fig}[1]{Figure~\ref{fig:#1}}

\newcommand{\tab}[1]{Table~\ref{tab:#1}}
\newcommand{\be}{\begin{equation}}
\newcommand{\ee}{\end{equation}}
\newcommand{\ba}{\begin{align}}
\newcommand{\ea}{\end{align}}
\newcommand{\bad}{\begin{equation} \begin{aligned}}
\newcommand{\ead}{\end{aligned} \end{equation}}
\newcommand{\bea}{\begin{eqnarray}}
\newcommand{\eea}{\end{eqnarray}}

\def\la{\langle}

\newcommand{\msun}{M_\odot}
\newcommand{\Msun}{M_\odot}

\newcommand{\ifm}[1]{\relax\ifmmode#1\else$\mathsurround=0pt #1$\fi}
\newcommand{\kms}{\ifmmode\,{\rm km}\,{\rm s}^{-1}\else km$\,$s$^{-1}$\fi}

\newcommand{\Mpc}{\,{\rm Mpc}}
\newcommand{\kpc}{\,{\rm kpc}}
\newcommand{\pc}{\,{\rm pc}}

\newcommand{\Gyr}{\,{\rm Gyr}}

\newcommand{\Myr}{\,{\rm Myr}}

\newcommand{\yr}{\,{\rm yr}}
\newcommand{\erg}{\,{\rm erg}}

\newcommand{\cmc}{\,{\rm cm}^{-3}}
\newcommand{\cms}{\,{\rm cm}^{-2}}

\newcommand{\ltsima}{$\; \buildrel < \over \sim \;$}
\newcommand{\lsim}{\lower.5ex\hbox{\ltsima}}
\newcommand{\gtsima}{$\; \buildrel > \over \sim \;$}
\newcommand{\gsim}{\lower.5ex\hbox{\gtsima}}
\newcommand{\prop}{\propto}

\def\omm{\Omega_{\rm m}}
\def\oml{\Omega_{\Lambda}}
\def\omb{\Omega_{\rm b}}

\def\Rv{R_{\rm v}}
\def\Vv{V_{\rm v}}

\def\Mg{M_{\rm g}}
\def\Ms{M_{\rm s}}
\def\Mselev{M_{\rm s,11}}
\def\Re{R_{\rm e}}

\def\Sig1{\Sigma_1}

\def\Vsn{V_{\rm SN}}

\def\Rd{R_{\rm d}}

\def\Vrot{V_{\rm rot}}

\def\tac{t_{\rm ac}}

\def\brho{\bar\rho}

\def\Rd{R_{\rm d}}
\def\Hd{H_{\rm d}}

\def\drhos{\rho_{\rm sfr}}

\def\rhog{\rho_{\rm g}}

\def\epsf{\epsilon_{\rm ff}}
\def\eps2{\epsilon_{-2}}

\def\tff{t_{\rm ff}}

\def\tam{t_{\rm mer}}

\def\tac{t_{\rm ac}}
\def\torb{t_{\rm orb}}
\def\tv{t_{\rm v}}
\def\Mv{M_{\rm v}}
\def\Mvd{\dot{M}_{\rm v}}
\def\M11{M_{\rm v,11}}

\def\Md{M_{\rm b}}
\def\Mdd{\dot{M}_{\rm b}}
\def\Vd{V_{\rm d}}

\def\fs{f_{\rm sb}}
\def\fsh{f_{\rm sb,0.5}}
\def\fb{f_{\rm b}}
\def\f16b{f_{\rm b,0.16}}
\def\svel{f_{{\rm sv},11,-2.5}}
\def\fsv{f_{\rm sv}}
\def\sv25{f_{{\rm sv},-2.5}}
\def\epsf{\epsilon_{\rm ff}}
\def\jd{j_{\rm b}}
\def\jin{j_{\rm in}}

\title[Mass threshold for discs]
{A Mass Threshold for Galactic Gas Discs by Spin Flips}
\author[Dekel et al.]
{\parbox[t]{\textwidth}
{Avishai Dekel$^{1,2}$\thanks{E-mail: dekel@huji.ac.il},
Omri Ginzburg$^1$,
Fangzhou Jiang$^1$,
Jonathan Freundlich$^1$,
Sharon Lapiner$^1$,
Daniel Ceverino$^3$,
Joel Primack$^4$
}
\\ \\
$^1$Racah Institute of Physics, The Hebrew University, Jerusalem 91904 Israel\\
$^2$SCIPP, University of California, Santa Cruz, CA 95064, USA\\ 
$^3$Departamento de Fisica Teorica, Facultad de Ciencias, Universidad Autonoma
de Madrid, Cantoblanco, 28049 Madrid, Spain\\ 
$^4$Physics Department, University of California, Santa Cruz, 1156 High Street, Santa Cruz, CA 95064, USA
}

\begin{document}

\large

\pagerange{\pageref{firstpage}--\pageref{lastpage}} \pubyear{2002}

\maketitle

\label{firstpage}

\begin{abstract}
We predict, analytically and by simulations, that gas discs tend to 
survive only in haloes above a threshold mass $\sim\!2\! \times\! 10^{11}\msun$ 
(stellar mass $\sim\!10^9\msun$), with only a weak redshift dependence.  
At lower masses, the disc spins typically flip in less than an orbital time 
due to mergers associated with a change in the pattern of the feeding 
cosmic-web streams.  This threshold arises from the halo merger rate
when accounting for the mass dependence of the 
ratio of galactic baryons and halo mass.
Above the threshold, wet compactions lead to massive central nuggets 
that allow the longevity of extended clumpy gas rings. 
Supernova feedback has a major role in disrupting discs below the
critical mass, by driving the
stellar-to-halo mass ratio that affects the merger rate, by stirring up
turbulence and suppressing high-angular-momentum gas supply,
and by confining major compactions to the critical mass.
Our predictions seem consistent with current observed fractions of gas discs, 
to be explored by future observations that will resolve galaxies below
$10^9\msun$ at high redshifts, e.g. by JWST.
\end{abstract}


%%%%%%%%%%%%%%%%%%%%%%%%%%%
\section{Introduction}
\label{sec:intro}

% dry mergers to elliptical -- not here
Galaxies tend to appear as discs, triaxial spheroids and combinations of the 
two, as well as in irregular morphologies.
For massive galaxies, 
\adb{e.g., in dark-matter haloes of $\sim\!10^{12}\msun$ and above,}
the fraction of quenched spheroids is known to be
growing with cosmological time and with mass at the expense of discs,
consistent with the common wisdom that discs tend to become spheroids
due to dry mergers. We do not address this well-studied phenomenon here, 
but rather focus on the survival versus disruption of gas discs,
especially at lower masses.

\smallskip % wish to find range of discs
On intermediate and small mass scales, 
\adb{in haloes of $10^{11}\msun$ and below, say,}
where the shape and kinematics are harder
to resolve, the observational situation is less clear.
\adb{Partly relevant observational attempts are discussed in \se{obs},
by \citet{cortese14} and \citet{elbadry18b} at low redshift based on the SAMI 
survey,
and by \citet{kassin12} and \citet{simons17} out to $z\!=\!2.4$.
Lower masses at higher redshifts are hopefully to be explored with future 
telescopes such as JWST.}
\adb{Earlier cosmological simulations, discussed in \se{other_sims},
reveal that at low redshift discs do not tend to survive in low-mass galaxies
\citep{elbadry18a,wheeler19}}.
One wishes to find out the range of disc viability
in terms of cosmological epoch and mass.
In particular, can star-forming gas discs form and survive at high redshifts?
Can they form and survive at low masses?

\smallskip % expect a redshift threshold
Discs are fragile. They are expected not to survive under intense mass
accretion including mergers, especially major mergers,
if the orbital angular momentum and spins of the
incoming objects are not aligned with the angular momentum of the disc.
A naive estimate guided by the merger rate of dark-matter haloes would lead to 
the apparent expectation that discs of all masses had hard time surviving at 
high redshifts. 
This is because, in the Einstein-deSitter cosmology regime (practically $z>1$),
the characteristic time for the occurance of halo mergers is 
$t_{\rm mer} \prop (1+z)^{-5/2}$ \citep{neistein08_m,dekel13},
while the galaxy orbital time is proportional to the Hubble time,
$t_{\rm orb} \prop (1+z)^{-3/2}$.
The ratio, $t_{\rm mer}/t_{\rm orb} \prop (1+z)^{-1}$, is thus smaller at higher
redshifts, indicating that there is a critical redshift above which major 
mergers occur in every disc orbital time and thus do not allow a long-lived 
disc.

\smallskip % mass dependence in sims
Contrary to this naive expectation, using below zoom-in cosmological 
simulations to explore the relative abundance of discs in the plane of halo 
mass $\Mv$ and redshift $z$, we find instead a threshold mass for disc 
viability, at $\Mv \sim 2\times 10^{11}\msun$,
roughly independent of redshift.

\smallskip  % mass dependence of flips
We show that this is actually expected when recalling that
the ratio of baryonic mass of a galaxy (stars plus cold gas) 
to its total halo mass
is increasing systematically with mass and with redshift\footnote{Note that
this redshift dependence is true for the baryonic mass in the galaxy, 
while for the stellar mass the ratio
is roughly constant with $z$ in the range $z\!=\!0\!-\!4$
\citep{moster10,behroozi13,behroozi19}, or even decreasing with growing $z$
\citep{rodriguez17}.}.
While the external inflow (and merger) rate of mass and angular-momentum 
that could damage a disc is primarily determined by the total halo mass, 
the angular momentum of the existing central galaxy that it is affecting 
is increasing with its baryonic mass. Therefore the damaging {\it relative} 
change in angular momentum is expected to be larger for lower-mass galaxies 
and at lower redshifts.  This should introduce a strong mass dependence in 
the disc survivability and weaken its redshift dependence.
Indeed, when evaluating analytically
the disc survivability under angular-momentum flips due to mergers,
as a function of mass and redshift, we predict
a threshold halo mass at $\Mv\! \sim\! 2\times 10^{11}\msun$
with only little sensitivity to redshift, consistent with the threshold
seen in the simulations.

\smallskip % compaction-driven ring
\adb{Another phenomenon that points to a characteristic mass of a similar value
is the process of wet compaction into a blue nugget, sometimes driven by
mergers and in other times by other mechanisms such as counter-rotating streams
\citep{zolotov15,tacchella16_ms,tacchella16_prof,tomassetti16,dekel19_gold}.}
We address below, and in more detail in \citet{dekel20_ring},
how the compaction process may have a role in the transition from disc 
disruption to the survival of an extended gas disc, then ring,
at a similar threshold mass.
At low masses, pre compaction, a disc disruption is expected to be driven by  
inward transport of gas and clumps driven by violent disc instability 
and possibly also by torques exerted by the prolate inner system. 
At masses above the critical mass, namely post compaction, we show that an 
extended disc, actually a ring fed by cold streams of high angular momentum, 
is expected to survive under the influence of the massive, rounder, 
post-compaction bulge (or by a massive cuspy halo).
The ring survival is evaluated through the disruptive
torques exerted by a spiral structure \citep{dekel20_ring}.

\smallskip % SN
It is interesting to note that the origin of the threshold mass for extended
rotating discs or rings
by the above mechanisms may result in several ways from the robust mass 
dependence of supernova feedback, namely the upper limit for the mass of a 
halo within which supernova feedback can effectively heat up the gas or remove 
it and thus suppress star formation \citep{ds86}. 
Supernova feedback could directly disrupt discs of low masses by driving
excessive turbulence or by halting the supply of new high-angular-momentum gas.
Indirectly, the increase of baryon-to-halo mass ratio with mass, which 
determines the threshold due to merger-driven spin flips, is argued to arise 
from supernova feedback \citep{ds86,dw03}.
Furthermore, the characteristic mass for major compaction events is argued to 
also be due to the supernova feedback being effective in suppressing compaction
events in galaxies of lower masses \citep{dekel19_gold}.

\smallskip
The paper is organized as follows.
In \se{sims} we briefly present the VELA simulations, 
which are described in Appendix \se{app_vela},
and present the threshold mass for discs as seen in the simulations.  
In \se{flip_sim} we show that the disc disruption below the critical mass
is consistent with spin flips in an orbital timescale.
In \se{flip_toy} we work out an analytic evaluation of the threshold mass
for merger-driven spin flips.
In \se{compaction_ring} we briefly describe the possible effect of a compaction 
to a blue nugget on the longevity of extended discs and then
rings above the threshold mass, and refer to \citet{dekel20_ring} for more
details.
In \se{disc_SN} we discuss the key role of supernova feedback in the disruption
versus survival of discs.
In \se{disc_other} we discuss other simulation results and compare our
predictions to potentially relevant observations.
In \se{conc} we summarize our conclusions.

%%%%%%%%%%%%%%%%%%%%%%%%%%%%%%%%%% 2
\section{Mass Threshold in Simulations}
\label{sec:sims}

%===================   %2.a
\subsection{The VELA simulations}
\label{sec:vela}

\begin{figure*} % 1
\centering
\includegraphics[width=0.48\textwidth]{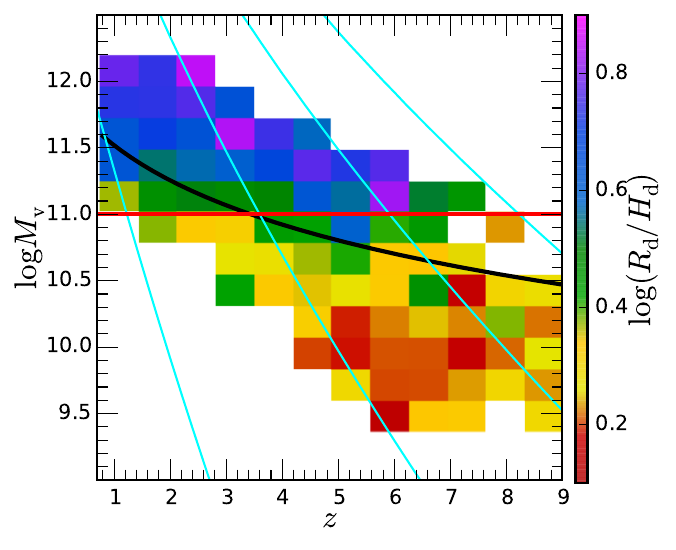}
%\qquad
%\quad
\includegraphics[width=0.48\textwidth]{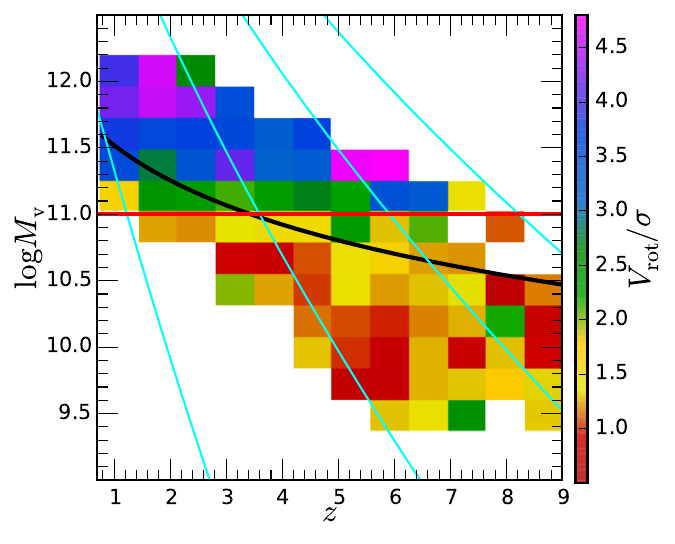}
%\special{psfile="figs/fig1a.eps" %log_rdhd_hist2d_mv_z.eps" 
%         hscale=80 vscale=80 hoffset=-120 voffset=-215}
%\special{psfile="figs/fig1b.eps" %vsig_hist2d_mv_z.eps" 
%         hscale=80 vscale=80 hoffset=140 voffset=-215}
\caption{
Discs versus non-discs in the VELA simulations.
The disc measures $\Rd/\Hd$ and $\Vrot/\sigma$ (color) are
averaged within bins in the mass-redshift plane.
We see a mass threshold for disc viability at $\Mv\!\sim\!10^{11}\msun$
(highlighted by red horizontal line),
with negligible redshift dependence, as predicted in \se{flip_toy}.
The two measures of disciness are consistent with each other.
The black curve refers to the upper limit for effective supernova feedback at
a virial velocity $\Vv\!=\!120\kms$ (\se{disc_SN}).
The cyan curves refer to the Press-Schechter $\nu\sigma$ peaks,
for $\nu=1,2,3,4$ from left to right, respectively.
}
\label{fig:disc_Mz_bins}
\end{figure*}

\begin{figure*} % 2
\centering
%\subfloat{\includegraphics[width=0.47\textwidth]{fig2a.png}}
\includegraphics[width=0.47\textwidth]{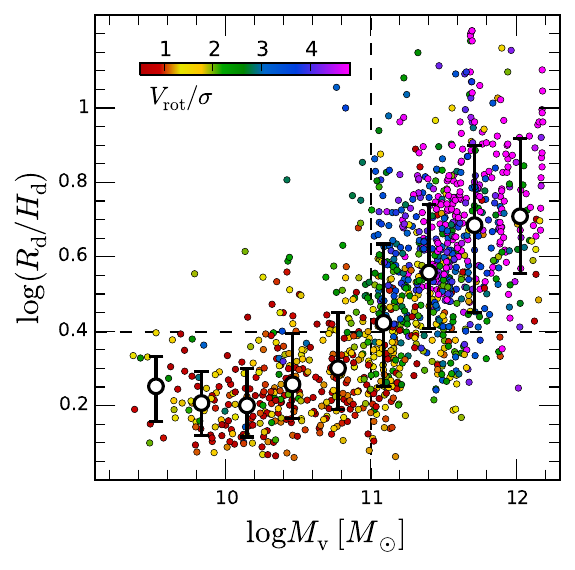}
\quad
\includegraphics[width=0.4476\textwidth]{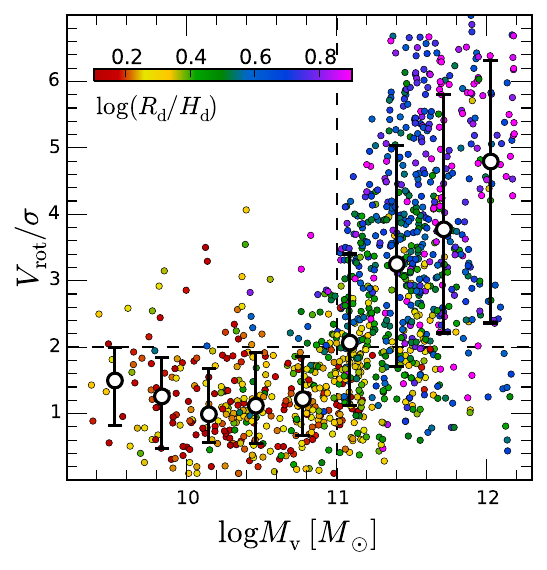}
%\vskip 8.5cm
%\special{psfile="figs/fig2a.eps" %log_rdhd_vs_log_mv_color_linear_vsig.eps"
%                 hscale=85 vscale=85 hoffset=-140 voffset=-215}
%\special{psfile="figs/fig2b.eps" %linear_vsig_vs_log_mv_color_log_rdhd.eps"
%                 hscale=85 vscale=85 hoffset=110 voffset=-215}
\caption{
From non-discs to discs in the VELA simulations.
Two measures of disciness, $\Rd/\Hd$ and $\Vrot/\sigma$,
as a function of halo mass $\Mv$ for all the snapshots of all the 34
evolving galaxies.
Each point refers to a snapshot, with the median and $1\sigma$ scatter
(16\% and 84\% percentiles) marked in bins of $\Mv$.
The color refers to the measure of disciness in the other panel.
The two measures of disciness are consistent with each other.
We see a marked transition from non-discs to discs at a threshold mass of
$\Mv \simeq 10^{11}\msun$, as predicted.
}
\label{fig:disc_Mv}
\end{figure*}

\begin{figure*} % 3
\centering
\includegraphics[width=0.9\textwidth]{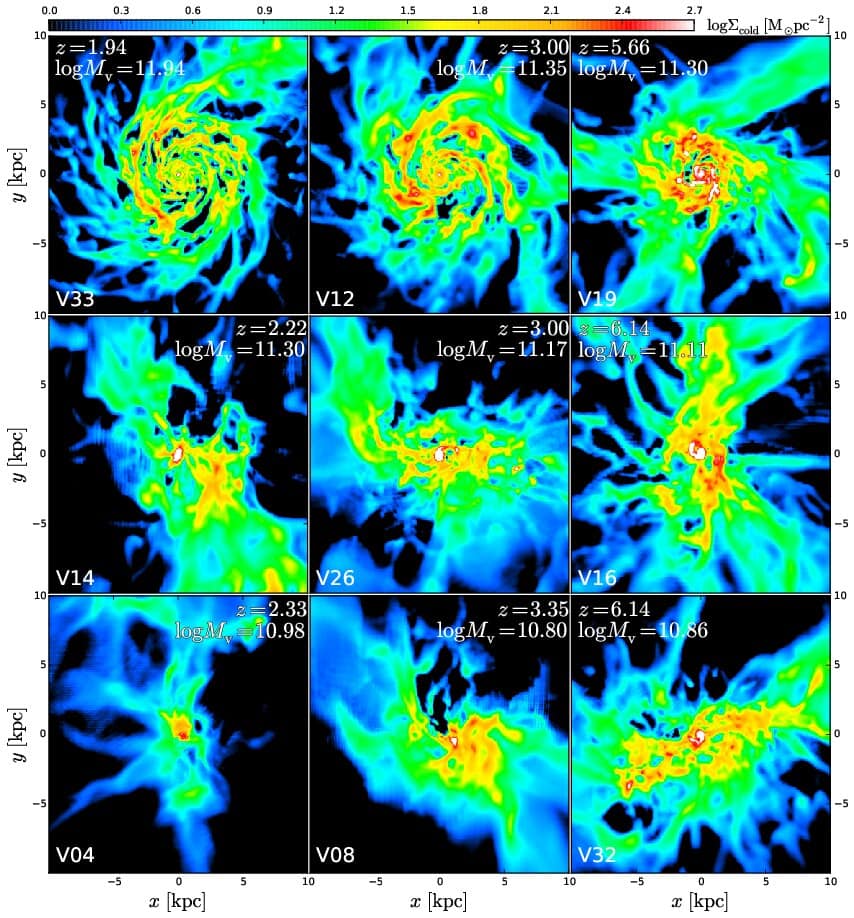}
%\vskip 16cm
%\special{psfile="figs/fig3.ps" %3x3_faceon_mosaic.ps"
%                 hscale=50 vscale=50   hoffset=40 voffset=0}
\caption{
Example VELA galaxies in different zones of the $\Mv-z$ plane.
Shown is projected gas density, face on with respect to the angular momentum.
The rows from top to bottom represent galaxies above, near and below
$\Mv=10^{11}\msun$,
while the columns from left to right represent low, intermediate and high
redshifts, as marked.
Stable, extended gas discs are seen only above the critical mass, while
the gas in galaxies below the critical mass is amorphous, 
showing a compact blue nugget near the critical mass.
}
\label{fig:Mv-z_face}
\end{figure*}

% VELA
In parallel to our analytic modeling, 
we utilize a suite of 34 VELA zoom-in hydro-cosmological simulations with
high resolution.
A few relevant characteristics are summarized here, while more details are
provided in \se{app_vela}, \tab{sample} and in several published papers
\citep[e.g.][]{ceverino14,zolotov15}, recently in \citet{dekel19_ks}.
The simulations are based on an Adaptive Refinement
Tree (ART) code \citep{krav97,ceverino09}. The suite consists of 34
galaxies that were evolved to $z\sim 1$, with a unique maximum spatial
resolution ranging from $17.5$ to $35 \pc$ at all times.
The dark-matter halo masses at $\!z=\!2$ range from $10^{11}$ to $10^{12}\msun$.
\adb{
The galaxies for the zoom-in runs were selected at $z\!=\!1$ from a 
lower-resolution dark-matter-only cosmological simulation, 
such that their dark-matter haloes are in the ball park of $\sim\!10^{12}\msun$
and they did not suffer a major merger (mass ratio larger than 1:4)
in the immediate vicinity of $z\!=\!1$. 
We know from the pre-run that this eliminated less than ten percent 
of the haloes in the simulated mass range, which, given the major mergers,
were likely to end up as massive ellipticals at low redshifts.} 
The halo mass selection also eliminated dwarf galaxies at $z<4$.

\smallskip % processes
Besides gravity and hydrodynamics, the code incorporates physical process
relevant for galaxy formation such as gas cooling by atomic hydrogen and
helium, metal and molecular hydrogen cooling, photoionization heating by the
UV background with partial self-shielding, star formation, stellar mass loss,
metal enrichment of the ISM and stellar feedback. Supernovae and stellar winds
are implemented by local injection of thermal energy 
\citep[as described in][]{ceverino09,cdb10,ceverino12}. 
Radiation-pressure stellar feedback is implemented at a moderate level, 
following \citet{dk13}, as described in \citet{ceverino14}.
In general, the stellar and supernova feedback as implemented in this suite 
is on the weak side of the range of feedback strengths in common cosmological 
simulations. 
\adb{This leads to a moderate underestimate of the gas fractions and 
a moderate overestimate of the stellar-to-halo mass ratios,
which may affect certain results quantitatively but is
not expected to hurt the predictive power of qualitative phenomena.}
\adb{No AGN feedback is incorporated, but in the mass range of interest
in the current analysis, halos near $\sim\!10^{11}\msun$ and below,  
its effects are expected to be small compared to those of stellar and supernova
feedback \citep[e.g.][]{dubois15,dekel19_gold}.
The unique high resolution of the VELA simulations is much more important for
the study of discs and low-mass galaxies.}

\smallskip
%Disk definition from Mandelker. Values in the table}
In the analysis of the simulations,
the disc plane and dimensions are determined iteratively, as detailed in
\citet{mandelker14}. In short, the disc axis is defined by the angular
momentum of cold gas ($T\! <\! 1.5 \times 10^{4}$K), 
which on average accounts for $\sim\! 97\%$ of the total gas mass in the disc.
The disc radius $\Rd$ is chosen to contain $85\%$ of the cold gas mass in the
galactic mid-plane out to $0.15\Rv$, where $\Rv$ is the halo virial radius.
The disc half-height $\Hd$ is defined to 
encompass $85\%$ of the cold gas mass in a thick cylinder where both the
radius and half-height equal $\Rd$. The values of the disc dimensions at
$z\!=\!2$ are listed in \tab{sample}.
The ratio $\Rd/\Hd$ is used below as one of the measures of gas disciness.

\smallskip % V and sigma
Another measure of disciness that is used in parallel is the kinematic ratio 
of rotation velocity to velocity dispersion $\Vrot/\sigma$. 
The rotation velocity and the velocity dispersion are computed by 
mass-weighted averaging over cells inside a cylinder whose minor axis is along 
the angular-momentum direction of the cold gas ($T\!<\!4\times 10^4$K) within a
sphere of radius $0.1\Rv$. The cylinder radius is $0.1\Rv$ and its half-height
is $0.25 \Re$, where $\Re$ is the cold-gas half-mass radius
(more details in Kretchmer et al., in prep.).
\adb{
The radial velocity dispersion is measured with respect to the mean radial
velocity within radial bins, 
and then averaged mass-weighetd within the cylinder.
The exact way the disciness is measured turns out not to have a significant 
effect on the relative measure of disciness between galaxies, which is our main
concern here.}

\begin{figure*} % 4 a,b
\centering
\includegraphics[width=0.48\textwidth]{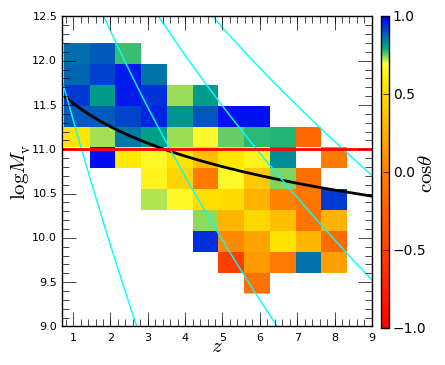}
%\quad
\includegraphics[width=0.48\textwidth]{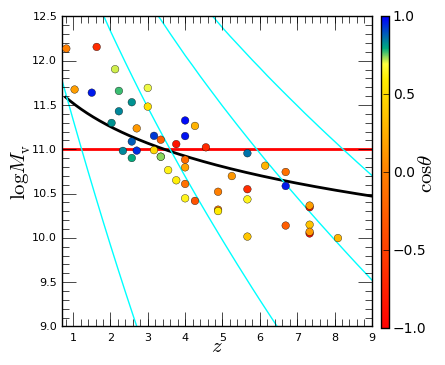}
%\vskip 7.2cm
%\special{psfile="figs/fig4a.eps" %cost_hist2d_mv_z.eps" 
%          hscale=80 vscale=80 hoffset=-120 voffset=-215}
%\special{psfile="figs/fig4b.eps" %cost_mv_z_merger_pts.eps" 
%         hscale=80 vscale=80 hoffset=140 voffset=-215}
\caption{
Disc flips in the VELA simulations, as measured by
the angle between the gas spins in two output times separated by a disc orbital
time (color), in the mass-redshift plane.
{\bf Left:}
The cosine of the angle averaged within bins in the mass-redshift plane,
showing that flips of $\theta > 45^\circ$ dominate below a mass of
$\Mv\!\sim\!10^{11}\msun$, while alignments dominate above this threshold,
with a negligible dependence on redshift, as predicted.
{\bf Right:}
The cosine of the angle (color) for snapshots where major mergers occur within
an orbital time, showing a tendency for disc flips to be associated with major
mergers for $\Mv\!<\!10^{11}\msun$.
The black curves show the $\nu\sigma$ Press-Schechter mass for
$\nu\!=\!1\!-\!4$ from left to right.
We see that disc flips tend to occur for mergers of high-sigma peaks,
$\nu \geq 2$, which are associated with mergers of nodes of the cosmic web.
The reference curves for SN feedback and $\nu\sigma$ peaks are
shown as in \fig{disc_Mz_bins}
}
\label{fig:flip_Mz}
\end{figure*}

%====================% 2.b
\subsection{Mass threshold for discs in the simulations}
\label{sec:threshold_sims}

We use the simulations to explore how discs and non-discs populate the
$\Mv\!-\!z$ plane.
\Fig{disc_Mz_bins} shows the two measures of gas disciness, 
$\Rd/\Hd$ and $\Vrot/\sigma$, in this plane. 
The ratios $\Rd/\Hd$ and $\Vrot/\sigma$ (color) are averaged over all the 
simulation snapshots in bins of $\Mv$ and $z$.
We see a systematic gradient of ``disciness" with mass,
and a division between the zones of non-disc and disc 
dominance at a critical halo mass of 
$\Mv\!\simeq\!(1\!-\!2)\times 10^{11}\msun$, 
with no significant redshift dependence.
This is predicted by simple analytic model below, contrary to the naive
expectation mentioned in the Introduction of a redshift threshold based 
on halo merger rates. 

\smallskip
Focusing on the mass dependence,
\fig{disc_Mv} shows the distribution of the two measures of gas disciness
as a function of halo mass.
The symbols correspond to all the $z\!<\!9$
snapshots of all the 34 simulated galaxies
as they evolve in time and grow in mass.
We again see a clear transition from non-disc to disc dominance
at about $\Mv \simeq 10^{11}\msun$,
separated at $\Rd/\Hd\!\simeq\!2.5$ and $\Vrot/\sigma\!\simeq\!2$.

\begin{figure*} %5
\centering
\includegraphics[width=0.9\textwidth]{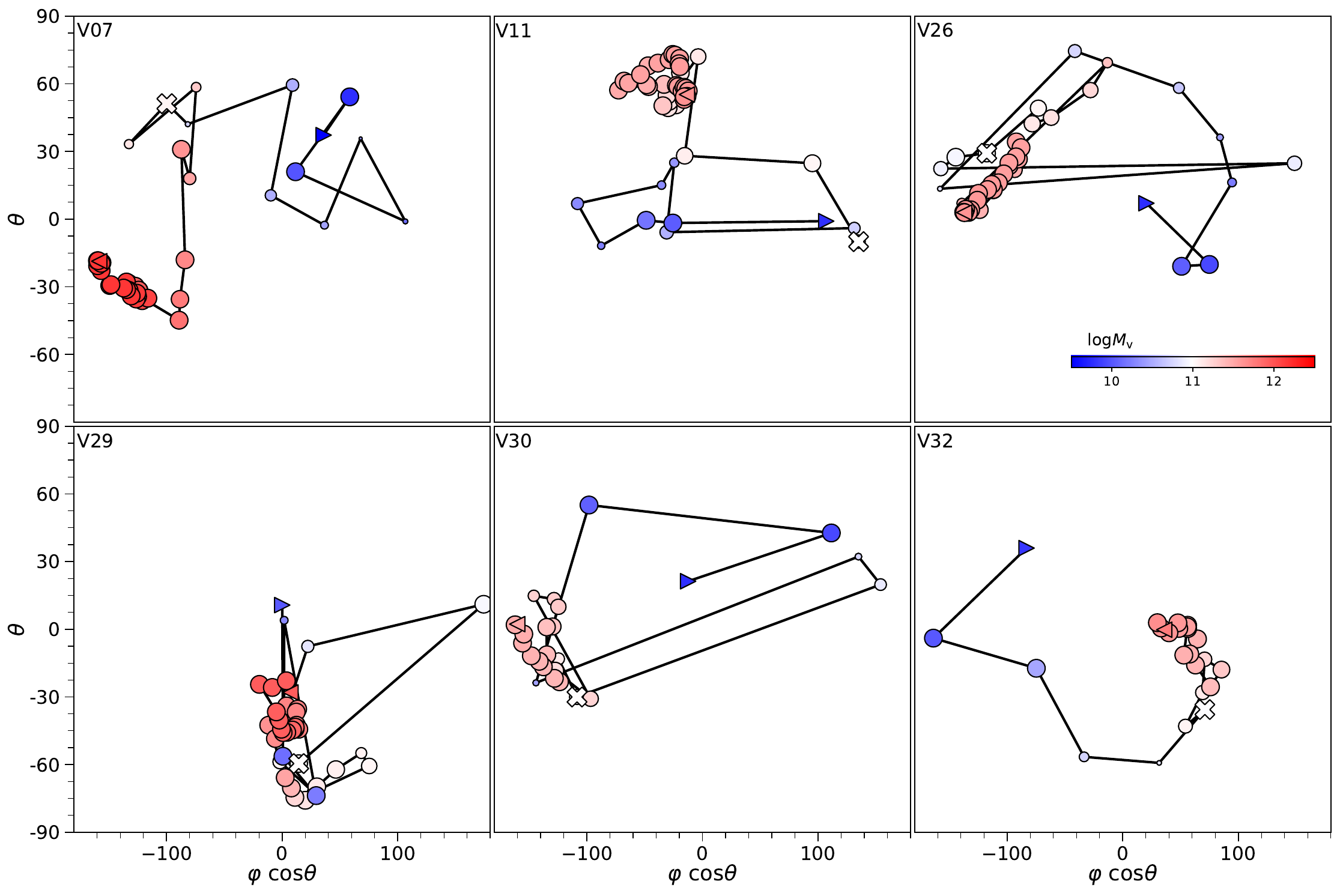}
%\vskip 11.0cm
%\special{psfile="figs/fig5.eps" %AMtrack_mosaic.eps" 
%         hscale=44 vscale=44 hoffset=105 voffset=-20}
\caption{
Evolution tracks on the sky of spin direction and magnitude (symbol size)
for six VELA galaxies.
The angles $\theta$ and $\phi$ are the standard spherical coordinates in
a randomly oriented frame, such that the line intervals fairly
represent the angles between the spins in successive snapshots.
The color indicates halo mass $\Mv$ as the galaxy is growing in time.
Blue and red triangles mark the earliest and latest times, typically
$z\!\sim\!8$ and $1$, when the galaxies are of the lowest and highest mass,
respectively.
The white cross marks $\sim\!10^{11}\msun$.
The galaxy orbital time ranges from less than a time step to a couple of time
steps from the initial to the final snapshots.
We see frequent spin flips as long as $\Mv\!<\!10^{11}\msun$ (blue),
relaxing to a stable spin direction and amplitude at $\Mv\!>\!10^{11}\msun$
(red), post compaction.
}
\label{fig:spin_sky}
\end{figure*}

\begin{figure} % 6
\centering
\includegraphics[width=0.48\textwidth]{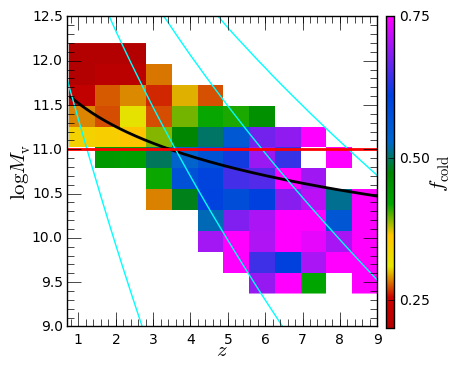}
%\vskip 7.5cm
%\special{psfile="figs/fig6.eps" %fcold_hist2d_mv_z_015rvir.eps" 
%        hscale=80 vscale=80 hoffset=-130 voffset=-215}
\caption{
%\adr{consider fgas instead of fcold}
Cold mass fractions with respect to the total baryons within $0.15\Rv$,
averaged within bins in the $\Mv\!-\!z$ plane.
The cold mass includes cold gas and young stars, with the latter contributing
about a third of the mass.
There is a gradient with virial mass, crossing $\sim\! 0.5$ near the critical
mass, indicating that the mergers in low mass galaxies tend to be wet,
and therefore more disruptive to the gas discs.
}
\label{fig:gas_frac}
\end{figure}

\smallskip
To illustrate the distribution of disciness in the $\Mv\!-\!z$ plane,
\fig{Mv-z_face} shows images of face-on projected gas density in example VELA
galaxies that fall in the different zones of the $\Mv\!-\!z$ plane. 
The corresponding edge-on images are shown in \fig{Mv-z_edge}. % xxx
Indeed, we see that the gas in galaxies of $\Mv\!>\!10^{11}\msun$ is typically 
in an extended rotating disc. In contrast, the gas in galaxies of 
$\Mv\!<\!10^{11}\msun$ is typically irregular and undergoing mergers. 
In between, the gas in galaxies of $\Mv\!\sim\! 10^{11}\msun$ is typically in
a compact configuration - a blue nugget generated by wet compaction.
\adb{More examples of non-discs at low masses and high redshifts are shown in
\fig{non-discs}, referring to the pre-compaction phase or during the 
compaction process, showing a disturbed morphology and wet mergers in progress.}

%%%%%%%%%%%%%%%%%%%%%%%%%%%%%%%%%%%% 3 
\section{Spin flips in the simulations}
\label{sec:flip_sim}

%===========================
\subsection{Spin flips below the critical mass}

In order to see whether the threshold mass for discs at
$\Mv\!\sim\!2\times 10^{11}\msun$ indeed correlates with the rate of spin flips
in a galaxy orbital time,
the left panel of \fig{flip_Mz} shows the average change in the spin direction
over an orbital time, $\cos\theta$, in bins of $\Mv$ and $z$.
We indeed see a strong mass dependence with no apparent redshift dependence.
For $\Mv\!<\!10^{11}\msun$, the tilt angle is typically large,
$\cos\theta\!<\!0.7$ ($\theta\!>\!45^\circ$), which we can consider
``a flip".
In contrast, for $\Mv\!>\!2\times 10^{11}\msun$, the tilt is small,
$\cos\theta\!>\!0.85$ ($\theta\!<\!30^\circ$), namely the disc orientation is
stable over an orbital time.

\smallskip
To explore the association of major mergers with spin flips, the right panel of
\fig{flip_Mz} shows all major mergers in the $\Mv\!-\!z$ plane, with the color
referring to the tilt $\cos\theta$ over an orbital time about the merger.
We see that the mergers of galaxies in haloes of mass below the threshold
tend to be associated with flips, while the mergers of more massive haloes tend
to have a stable spin direction despite the merger.
This is with the exception of three massive outliers, possibly affected by
the fact that these mergers occur near the final time of the simulation making
it hard to assess flips in a full orbital time.

\smallskip % high-sigma
It is interesting to note in the right panel of \fig{flip_Mz} the
tendency of high-sigma peaks,  $\nu\!\geq\!2$
(near and to the right of the second black curve from the left),
to flip in association with major mergers. This is consistent with the tendency
of non-discs to populate the high-sigma peaks regime in \fig{disc_Mz_bins}.
It is along the lines of the
notion that mergers of high-sigma peaks, at the nodes of the cosmic web, are
associated with a change in the pattern of streams that feed the galaxy in the
node, including filament and sheet mergers,
which is in turn associated with a change in the angular momentum that is
brought by these streams to the galaxy \citep{danovich15}.
Mergers of low-sigma peaks, occurring at high masses and low redshifts,
tend to occur instead along cosmic-web filaments. These are less likely to
be associated with a change in stream pattern that causes a spin flip.

\smallskip
\Fig{spin_sky} makes the case for a mass threshold for spin flips even more
explicit. It shows the evolution of gas spin of each of six example VELA
galaxies. In each output time, the position of the symbol marks the spin
direction (with arbitrary zero-points for the spin direction in spherical
coordinates)
and its size indicates the magnitude of the angular momentum.
The starting and ending snapshots, typically at $z\!\sim\!8$ and $z\!\sim\!1$,
are marked by a blue and a red triangle. The halo mass is indicated by color,
with the snapshot closest to $10^{11}\msun$ marked by a white cross.
In all six cases we
clearly see big flips between output times, which are comparable to the
orbital times, as long as $\Mv\!<\!10^{11}\msun$ (blue to white symbols),
and a stable spin when $\Mv\!>\!2\times 10^{11}\msun$ (pink to red symbols).

\smallskip
\adb{The role of flips at different masses could in principle be either
due to a mass dependence of the merger rate or of the typical mass ratios 
in the mergers.
The latter is unlikely given the weak mass dependence of halo merger rates
as derived using the EPS formalism \citep{neistein08_m}}
%We conclude that the analytic estimate for a critical mass for
%spin flips in an orbital time, \equ{mass_threshold}, is confirmed in the
%simulations.
In \se{flip_toy} below we evaluate analytically the threshold mass for 
merger-driven spin flips 
\adb{based on the meregr rate, and recover the threshold mass 
%and in agreement with what we 
extracted from the simulations.}
%This implies 
We learn that merger-driven flips can be the reason for the disruption
of discs below the critical mass.

%==========================
\subsection{Wet mergers}

The disruption by each merger-driven flip can be more effective when the
mergers are more gas rich.
\Fig{gas_frac} shows the cold fraction with respect
to the total baryons within the larger disc radius, $0.15\Rv$, averaged in bins
within the $\Mv\!-\!z$ plane.
The young stars contribute about a third of the cold mass.
\adb{This fraction is not expected to vary drastically with $\Mv$ or $z$ 
as long as the same Kennicutt-Schmidt relation between SFR and gas density is 
valid, but it implies that uncertainties of order a factor of two are expected.}
We see a strong gradient with virial mass, rising to above $50\%$ below the
critical mass for merger-driven disc flips and for major compaction events
(see below).
We learn that below the critical mass the major mergers are not only more
frequent per orbital time (e.g., \fig{flip_Mz}),
but they are more gas rich, and therefore more disruptive for the gas discs.
This emphasizes further the transition from non-discs to discs near
the critical mass.

\smallskip
We discuss below how other mechanisms
can contribute to the disruption of discs below the critical mass.
These include the effects of supernova feedback, \se{disc_SN},
and the inward mass transport within violently unstable discs (VDI),
which may be
effective below a similar critical mass that refers to compaction events,
to be discussed next.

%%%%%%%%%%%%%%%%%%%%%%%%%%%%%%%%% 4
\section{Derivation of the threshold mass}
\label{sec:flip_toy}

%===================== 
\subsection{Flips in an orbital time}
\label{sec:flip_orb}

We expect a disc to be disrupted once the direction of its angular-momentum 
(AM) is changed significantly during a disc orbital time.
We assume that such a flip occurs in a major merger of high-sigma nodes of the
cosmic web, which is associated with a pattern change of the cosmic-web
streams that feed the galaxy with mass and angular momentum 
\citep[][see below]{danovich15}.
The characteristic time $\tam$ between mergers of baryon mass ratio $r$ 
($\leq\! 1$) can be related to the baryon accretion timescale by 
\be
\tam \simeq \beta (r)\, \frac{\Md}{\Mdd}\, , 
\label{eq:tam0}
\ee
where $\Md$ is the galaxy baryonic mass, 
$\Mdd$ is the total baryonic accretion rate, 
and $\beta(r)$ is a factor of order unity for major mergers of
$r\!\sim\!1/3$ (more in \se{flip_merger}).

\smallskip
We write
\be
\tam \simeq \frac{\Md}{\Mdd} 
= \frac{\Ms}{\Mv}\, \frac{\Md}{\Ms}\, \frac{\Mv}{\fb \Mvd} \, ,
\label{eq:Mdd}
\ee
where $\Ms$ is the stellar mass and $\Mv$ is the virial mass.
We made here the assumption that the baryonic accretion rate relates to the 
total accretion rate via 
\be
\Mdd = \fb\, \Mvd \, ,
\ee
where $\fb\!=\! 0.16\, \f16b$ is the universal baryonic fraction.
This is supported by simulations \citep{dekel13}, and by a match to the
observed SFR on the main sequence as a function of redshift 
\citep{dekel09,dm14,rodriguez16}\footnote{This is as long as 
$\Mv \geq 10^{9.5}\msun$ 
such that gas accretion is not prevented by the UV background and the escape 
velocity from the halo \citep{sd03,okamoto08}.}.

\smallskip
In the third term of \equ{Mdd}, 
$\Mv/\Mvd$ is the average timescale for total accretion into the halo. 
It can be derived analytically in simple terms \citep{dekel13}
using the self-invariant Press-Schechter time variable (the inverse
of the linear growing mode of density fluctuations),
in good agreement with simulations \citep{dekel13}.
In the EdS regime ($z>1$), the approximation is 
\be
\tac = \frac{\Mv}{\Mvd} = \tau\,\Gyr\, (1+z)^{-5/2}\, \M11^{-0.1} \, ,
\label{eq:tac}
\ee
where $\Mv = 10^{11}\msun\, \M11$
and the weak mass dependence with the small power index $-0.1$
is an estimate for the LCDM power spectrum in the vicinity of
$\Mv \sim 10^{11}\msun$.
The value of $\tau$ as derived from the simulations is $\sim\!30$, 
so we denote below $\tau = 30\, \tau_{30}$.

\smallskip
The first term in \equ{Mdd} is the stellar-to-halo mass ratio, which can be 
estimated for 
LCDM haloes via abundance matching to observed galaxies. 
We term 
$\fsv\!=\!\Ms/\Mv$ and denote $\fsv\!=\!10^{-2.5} \sv25$.
The second term in \equ{Mdd} is the inverse of the stellar fraction in the 
galaxy with respect to the total baryons, which we term
$\fs=\Ms/\Md$ and denote $\fs \!=\! 0.5\,\fsh$.
We obtain from \equ{tam0}, \equ{Mdd} and \equ{tac}
\bea
\tam \simeq& 1.1\Gyr\, \sv25\,\fsh^{-1}\,
\tau_{30}\, \f16b^{-1}\, \beta \nonumber \\ 
          &\times (1+z)^{-5/2}\, \M11^{-0.1}\, .
\label{eq:tam2}
\eea
Recall that $\tau_{30}$, $\f16b^{-1}$ and $\beta$ are all parameters of order
unity, while $\fsv$ and $\fs$ depend on $z$ and/or $M$.

%----------------------
%\subsubsection{Flips in an orbital time}

\smallskip
In comparison, the disc orbital time is
\be
\torb = 2\pi\,\frac{\Rd}{\Vd}
      = 2\pi \frac{\lambda\,\Rv}{\Vv} \, ,
\ee
where $\Vv$ and $\Vd$ are the virial velocity and disc rotation velocity
respectively, and 
where $\lambda = \Rd/\Rv$ is the contraction factor from the halo
to the gas disc\footnote{This is sometimes identified with the halo spin
parameter but it has been shown not to correlate with the spin parameter
on a halo-by-halo basis \citep{jiang19_spin}.}.
We write $\lambda = 0.08\,\lambda_{0.08}$, where $\lambda \sim 0.08$ is derived
for the gas from the VELA simulations.
The virial crossing time in the EdS regime is \citep{dekel13}
\be
\tv = \frac{\Rv}{\Vv} =0.14\, t_{\rm hubble} \simeq 2.45\Gyr\, (1+z)^{-3/2} \, .
\ee
This gives
\be
\torb \simeq 1.23\Gyr\,\lambda_{0.08}\, (1+z)^{-3/2} \, .
\label{eq:torb}
\ee 

\smallskip
The ratio of timescales, from \equ{tam2} and \equ{torb}, is 
\bea
\frac{\tam}{\torb}\simeq& 0.89\, \sv25\, \fsh^{-1}\,
\tau_{30}\, \lambda_{0.08}^{-1}\, 
\f16b^{-1}\, \beta  \nonumber \\ 
        &\times (1+z)^{-1}\, \M11^{-0.1}\, .
\label{eq:tamtorb1}
\eea

\smallskip
This may seem to indicate a redshift dependence, emerging from 
$\tac/\tv \prop (1+z)^{-1}$, which would have implied that disc disruption is 
more efficient at higher redshifts and is rather insensitive to mass.
This is misleading because $\fsv$ has a strong mass dependence
that becomes the dominant factor in $\tam/\torb$, while $\fs$ has a redshift 
dependence that roughly cancels the overall redshift dependence of 
$\tam/\torb$, as we see next.

%========================= 
\subsection{Baryon-to-halo ratio and the threshold mass}

From abundance matching of LCDM haloes to observed galaxies 
\citep[][Fig. 7]{behroozi19}, we read for $\Mv=10^{10-12}\msun$
\be
\fsv \equiv \frac{\Ms}{\Mv} \simeq 3\times 10^{-3}\, \svel\, \M11 \, , 
\label{eq:behroozi19}
\ee
where $\svel$ is the value of ${\fsv}_{-2.5}$ at $\Mv\!=\!10^{11}\msun$, and
$\svel\!\simeq\! 1$ with only little redshift dependence in the redshift range 
$z\!=\!1\!-\!5$. 
The ratio at a given mass is indicated to rise with $z$ at higher redshifts, 
but not so in other attempts to estimate this ratio 
\citep[e.g.][]{rodriguez17}, so we ignore here any variations with $z$.
An analytic argument based on energetics of supernova feedback predicts 
a qualitatively similar increase with mass as in \equ{behroozi19}
but with a slightly flatter mass dependence, 
$\Ms/\Mv \prop \Mv^{2/3}$ \citep[][see \se{SN-mass}]{ds86}. 
We adopt here the latest empirical abundance-matching estimate from
\equ{behroozi19} without a strong effect on the results. 

\smallskip
We determine $\fs$, the stellar mass fraction with respect to the baryonic
mass, from the data of \citet{tacconi18}. This is a compilation of more 
than 1400 molecular gas measurements in typical star-forming galaxies at 
$z\!=\!0\!-\!4$, including the PHIBSS \citep{tacconi13} and PHIBSS2 
\citep{freundlich19_PHIBSS2}
samples. The dependence of $\fs$ on mass and redshift
is evaluated through a simultaneous power-laws fit in terms of the variables
$(1+z)$, $\Ms$ and $\Delta {\rm MS}$, the deviation from the star-forming
main sequence ridge.
This is assuming that the variables can be separated and that the dependence on
each can be approximated by a power law.
The best fit obtained is 
\be
\fs = 1.27\, (1+z)^{-1}\, \Mselev^{0.14} \, ,
\label{eq:fs1}
\ee
where $\Ms = 10^{11}\msun\, \Mselev$.\footnote{Note 
that the value of $\fs$ here is properly smaller than unity only for $z$ 
safely above $0.27$.
Furthermore, this relation identifies the total gas mass with the molecular gas
mass, which is a sensible approximation only at $z$ well above $0.4$.
Thus, \equ{fs1} is a valid approximation only at redshifts of order unity 
and above.}

Combined with \equ{behroozi19}, the required $\fs^{-1}$ can be expressed in 
terms of $\Mv$ as
\be
\fs^{-1} = 0.88\, \svel^{-0.14}\, (1+z)\, \M11^{-0.28} \, .
\label{eq:fs2}
\ee
Similar exponents are obtained when carrying instead successive fits, as in
\citet{genzel15} and \citet{tacconi18}, 
although such fits may indicate that the variable
separation is not entirely justified, and they hint to a weak redshift
dependence of the exponents of $\Ms$ and $\Delta {\rm MS}$.

\smallskip
In fact, the redshift dependence of $\fs$ could be predicted using a bathtub
steady-state model, where mass conservation is assumed
as a balance between input and output of gas mass and stellar mass.
Combining equations 19, 24 and 26 in \citet{dm14}, we indeed obtain in the 
range $z \gg 1.5$ 
\be
\fs^{-1} \prop \frac{\torb}{\tac} \prop (1+z) \, ,
\label{eq:bath}
\ee 
assuming a constant star-formation-rate efficiency in a dynamical time, $\epsf$.
A zero-point of $\fs^{-1} \sim 0.8$ at $z=0$ is obtained for
$\epsf \sim 0.01$, and for $\tau \sim 30 \Gyr$ in the accretion time,
ignoring stellar mass in the accretion at very high redshift. 
%\adr{elaborate?}
One should note, however, that \equ{bath} is valid for a galaxy as it evolves,
so the redshift dependence at a fixed mass is slightly different.

\smallskip
We learn in passing by combining \equ{behroozi19} and \equ{fs2} that at a given
$z$
\be
\Md \prop \Mv^{1.72} \, .
\label{eq:MdMv}
\ee

\smallskip
Substituting \equ{behroozi19} and \equ{fs2} in \equ{tamtorb1} we obtain 
\be 
\frac{\tam}{\torb}\simeq 0.79\,
\tau_{30}\, \lambda_{0.08}^{-1}\, \svel^{0.86}\,
\f16b^{-1}\, \beta\, \M11^{0.62}\, .
\label{eq:tamtorb2}
\ee
Here all the parameters are assumed to be constants, 
independent of $\Mv$ and $z$.
We learn that the mass dependence of $\tam/\torb$ is pronounced, 
but the redshift dependence has vanished.
%\adr{consider using the analytic mass dependence instead}

%-------------------
%\subsubsection{Mass threshold for discs by spin flips}
%\label{sec:toy_flip_mass_threshold}
 
\smallskip
If we adopt $\tam/\torb\! =\! 1$ as the threshold for disc survival,
we finally obtain from \equ{tamtorb2} a redshift-independent virial mass 
threshold for disc survival,
\be
\M11 \simeq 1.36\, \tau_{30}^{-1.62}\, \lambda_{0.08}^{1.62}\, \svel^{-1.39}\,
\f16b^{1.62}\, \beta^{-1.62}\, .
\label{eq:mass_threshold} 
\ee
We expect all the parameters to be of order unity and independent of mass and
redshift.
We thus predict a threshold halo mass for disc survival at
$\Mv\! \sim\! (1\!-\!2)\!\times\! 10^{11}\msun$, corresponding to
$\Ms\! \sim\! (3\!-\!6)\! \times\! 10^8\msun\, \svel$.
This is indeed seen very convincingly in the simulations,
\fig{disc_Mz_bins} and \fig{disc_Mv}.

\smallskip
In summary, the relation $\tam\! \prop\! (1+z)^{-5/2}$ for haloes turns into
$\tam\! \prop\! \Mv^{0.62} (1+z)^{-3/2}$ when one incorporates the mass 
dependence of $\Ms/\Mv$ and the redshift and mass dependencies of $\Md/\Ms$.
The resultant redshift dependence balances the redshift dependence of $\torb$,
yielding a pure mass dependence for $\tam/\torb$, which is translated to
a mass threshold for disc survival with no redshift dependence. 

\smallskip
We note that a weak redshift dependence in the threshold mass
would appear if $\fsv=\Ms/\Mv$ actually
evolves with redshift, or if the log slope of the redshift dependence of $\fs$ 
in \equ{fs1} is somewhat different from unity, as may be
obtained based on a different method of linear-log multi-variable fit of the
scaling relation.
As for the former possibility, 
in the abundance-matching analysis of \citet{rodriguez17}, the zero point of 
$\fsv$ at a given halo mass is actually decreasing with increasing redshift 
in the redshift range $z\!=\!0\!-\!3$ and it remains constant at higher 
redshifts.
This behavior is different from the findings of \citet{behroozi19},
thus reflecting a significant uncertainty in the stellar-to-halo mass ratio
from abundance matching. 
A decrease of $\fsv$ with redshift
would introduce a negative redshift dependence in \equ{tamtorb2},
namely a positive redshift dependence in \equ{mass_threshold}, which
would result in a larger mass threshold at higher redshifts.  
This would imply a lower fraction of discs at higher redshifts at a fixed mass,
as indicated by certain observations at $z\!<\!2$ (see \se{obs}, \fig{simons}). 
Such a redshift dependence is not seen in the VELA simulations at higher
redshifts,
\fig{disc_Mz_bins}, because $\fsv$ does not vary significantly with redshift in 
these simulations, consistent with \citet{behroozi19} at $z\!<\!5$.

%============= 
\subsection{Mergers and spin flips} 
\label{sec:flip_merger}

Back to the merger timescale assumed in \equ{tam0},
it can be obtained by expressing the mass added by a merger in two ways,
\be
\tam\,\Mdd\, f_{r} = \tilde{r}\, \Md \, ,
\ee
where 
$f_{r}$ is the mass fraction in mergers with mass ratio $>\!r$ with respect
to the total accreted baryonic mass,
and $\tilde{r} \gsim r$ is the average ratio for such mergers that are of
a ratio $>\!r$.
This gives \equ{tam0} with $\beta(r) \!=\! {\tilde{r}}/{f_{r}}$.

\smallskip %beta
We can crudely estimate $\beta\!\sim\!1$ from major mergers of haloes,
which can be evaluated using the EPS formalism.
On one hand, $\tilde{r} \!\sim\! 2 r$ in the major-merger regime
\citep[based on][Fig.~7]{neistein08_m}.
On the other hand, the mass ratio for the baryonic masses, $r_{\rm b}$, is
smaller than the ratio for the total halo masses, $r_{\rm h}$,
because the baryon-to-halo mass ratio is an increasing function
of mass. We have $r_{\rm b}\! \prop\! r_{\rm h}^{1.72}$ 
in the relevant mass range (from \equ{behroozi19} and \equ{fs2}),
so $r_{\rm h}\! \sim\! 2 r_{\rm b}$ in the major-merger regime.
The corresponding values of $f_{r}$ differ by a factor of 2,
roughly balancing the difference between $r$ and $\tilde{r}$.
For haloes, using EPS,
we read from Fig.~7 of \citet{neistein08_m}
that $f_{r}\! \simeq\! 1/4$, $1/3$ and $1/2$ 
for $r\! =\! 1/2$, $1/3$ and $1/10$, respectively.
This gives $\beta \simeq 2$, $1$ and $0.2$ for these choices of $r$.
We adopt for major mergers $r\!=\!1/3$ and thus $\beta\! \simeq\! 1$
as our fiducial value.

\smallskip % cosmic web flip
For a high-sigma halo, we expect a major merger to be associated with a
significant change in the AM direction of the cosmological inflow.
This is because high-sigma haloes form in the nodes of the cosmic web,
and a coalescence of such nodes is likely associated with a
change of the pattern of streams that feed the galaxy with AM, 
including mergers of filaments and sheets of the cosmic web
\citep{danovich15}.
This is provided that the amplitude of the AM added during the time between
mergers, by the merger itself and the rest of the accretion,
is comparable to (or larger than) the galaxy AM.

\smallskip % Delta J
In order to estimate the relative amplitude of the AM added during time
$\tam$ to the galaxy, we write the AM as
$J \!=\! \Md \jd$ and the added AM rate as $\dot{J} \!=\! \Mdd \jin$,
were $\jd$ and $\jin$ are the specific angular momenta in the galaxy and in the
inflowing mass.
The incoming $j$ is expected to be larger than the $j$ of the mass that has
been accumulated earlier, because both the impact parameters of the
inflowing streams and their velocities scale with the virial radius
that is growing with time.
The ratio $\jin/\jd$ is estimated to be of order 2
based on the simulation results of \citet[][Fig. 1]{danovich15}.
A crude estimate in the same ballpark is obtained analytically
based on the halo mass growth, $\Mv \!\prop\! \exp(-\alpha z)$
\citep{dekel13}, and the assumption that the specific AM scales
with $\Rv\Vv\! \prop\! \Mv^{2/3} (1+z)^{-1/2}$, where $\Vv$ is the halo virial
velocity.
This implies that $\jin$ grows by a factor $\sim 2$ over the time interval
in which the mass roughly doubles. If $\jd$ of a galaxy of mass $\Md$ at the
time considered can be approximated by $\jin$ at the time when half the mass
was in place, then we get $\jin/\jd\! \sim\! 2$.
The added AM during time $\tam$ is thus
\be
\Delta J = \dot{J} \tam \sim 2 \beta J  \gsim J\, .
\ee
The magnitude of the added AM before the next merger is therefore expected to
be sufficient for
causing a flip in the galaxy AM direction for any sensible value of $r$.

%%%%%%%%%%%%%%%%%%%%%%%%%%%%%%%%%%%%% 5
\section{Post-Compaction Discs and Rings}
\label{sec:compaction_ring}

%================
\subsection{Mass transport \& compaction to blue nuggets}
\label{sec:compaction}

While we learned so far that discs above the threshold mass are not likely to 
suffer frequent merger-driven spin flips, violently unstable discs are 
themselves subject to inward mass transport which may in principle disrupt 
the discs from the inside even in the absence of major mergers.
In a disc undergoing violent disc instability (VDI),
the non-cylindrically symmetric density perturbations exert torques that
induce transport of AM outward.
AM conservation implies an associated mass transport inward, e.g., in terms of 
clump migration and gas inflow through the disc
\citep[e.g.][]{noguchi99}.
\citet{dsc09} estimated the corresponding evacuation time of a disc to be 
\be
t_{\rm inflow,disc} \sim \delta_{\rm d}^{-2}\, \torb\, ,
\quad \delta_{\rm d} \equiv \frac{M_{\rm d}}{M_{\rm tot}} \, ,
\label{eq:inflow_disc}
\ee
where the key variable $\delta_{\rm d}$ is the mass ratio of cold disc
to total mass within radius $r$.
%\be
%\delta_{\rm d} \equiv \frac{M_{\rm d}}{M_{\rm tot}} \, .
%\label{eq:deltad}
%\ee
This has been estimated either via the encounters between giant
clumps or based on the shear-driven mass-inflow rate of \citet{shakura73}
with the maximum AM flux density by \citet{gammie01}.
With $\delta_{\rm d}\! \lsim\! 1$, we thus expect an inward mass transport
within a few orbital times. 
This is contrary to our finding from the simulations 
that extended rings survive for long term once above the threshold mass
(\se{rings}).

\smallskip
An immediate suspect that could have a role in stabilizing discs
above the threshold mass is 
\adb{the process of wet compaction to a blue nugget,
which tends to occur near a similar characteristic mass \citep{zolotov15}.}
In the simulations, the major compaction event is typically
associated with the formation of an extended long-lived disc or ring. 
We argue that the post-compaction appearance of a massive bulge
is responsible for the transition from a disc to a ring
and for slowing down its inward mass transport.
This is addressed in detail in \citet{dekel20_ring} and summarized here.

\begin{figure*} % 7
\centering
\includegraphics[width=1.00\textwidth]{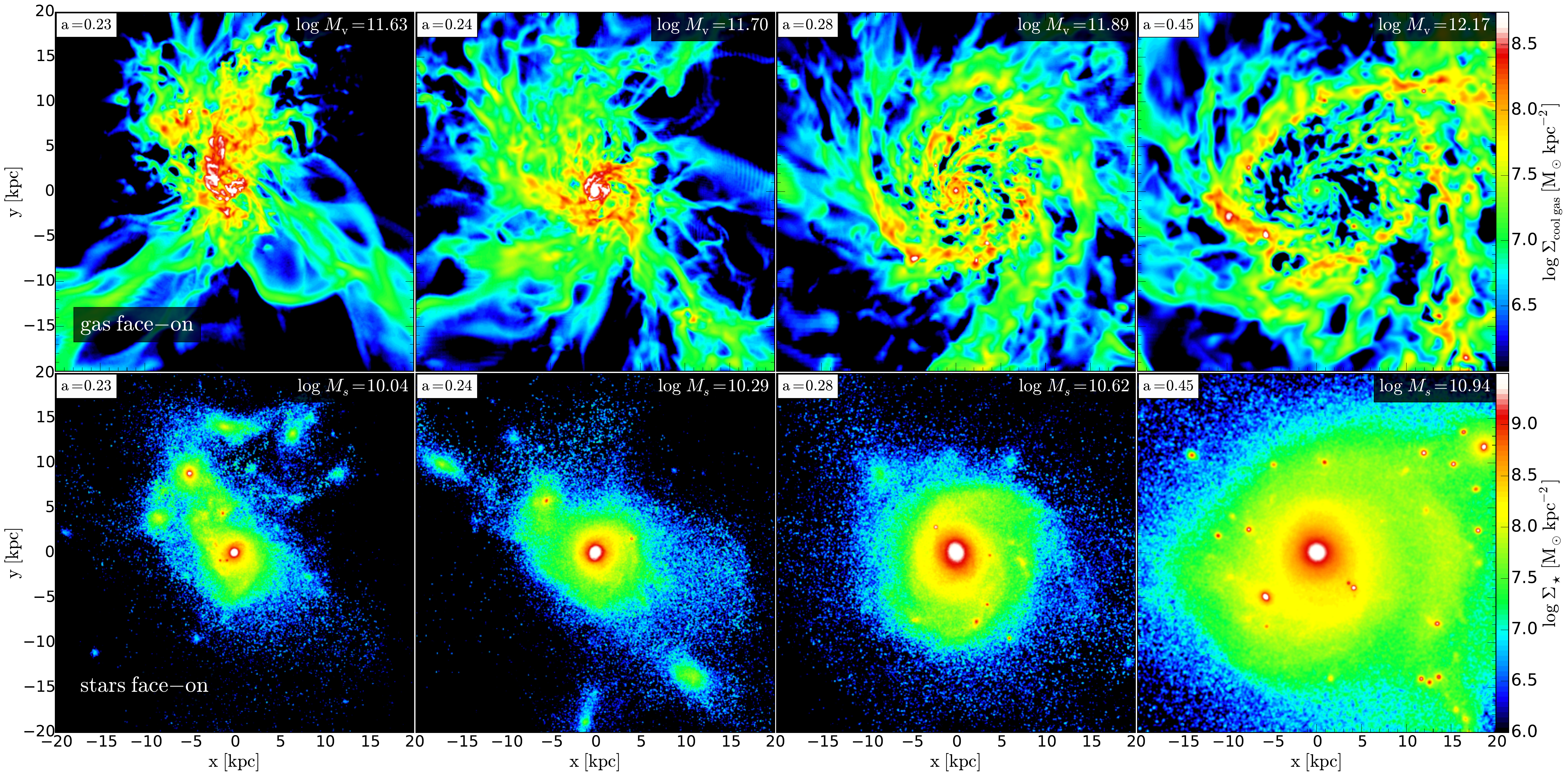}
%\vskip 9.0cm
%\special{psfile="figs/fig7.eps" %V07_faceon_seq8_cool-gas_stars_dpi30.eps"
%                 hscale=21.5 vscale=21.5   hoffset=185 voffset=45}
\caption{
Compaction to a blue nugget and post-compaction disc and ring.
Shown are the projected densities of gas (top) and stars (bottom)
in four phases (expansion factor $a=(1+z)^{-1}$ is marked)
during the evolution of one of the VELA simulated galaxies (V07).
The projections are face on with respect to the AM.
Top, from left to right.
First: during the compaction process.
Second: at the blue-nugget phase.
Third: post-compaction VDI disc.
Fourth: post-compaction, clumpy, long-lived ring, fed by incoming streams. 
Bottom: the stellar compact red nugget forms during and soon after the
compaction and the resulting bulge remains compact and grows massive 
thereafter.
}
\label{fig:mosaic_v07}
\end{figure*}

%========================== 5.1
%\subsection{Compaction to blue-nuggets}
%\label{sec:compaction}

\smallskip % compaction events
Cosmological simulations show that galaxies evolve through a dramatic
wet-compaction event, pronounced when the galaxy mass is near or 
above $\Mv\!\sim\! 10^{11.5}\msun$ ($\Ms\!\sim\! 10^{9.5}\msun$)
and in the redshift range $z\!=\!1\!-\!5$ when the gas fraction is high
\citep{zolotov15}.
This is gaseous contraction into a compact central
star-forming body within the inner $\sim\!1\kpc$, ``a blue nugget" (BN).
The gas consumption by star formation and ejection by
supernova feedback cause central gas depletion and
inside-out quenching of star-formation rate (SFR)
% at a roughly constant central stellar density 
\citep{tacchella16_prof}.
\Fig{mosaic_v07} below illustrates 
%through images of gas and stellar surface density 
the evolution through such a sequence of events 
in an example VELA galaxy.
The cartoon in Fig.~3 of \citet{dekel20_ring} shows the typical
evolution of gas mass, stellar mass and SFR within the inner kpc.
Figure 4 there shows evolution tracks of 
galaxies in the plane of specific SFR (sSFR) versus compactness as measured 
by the stellar surface density within $1\kpc$, $\Sigma_1$.
A compaction at a roughly constant sSFR turns into quenching at a constant
$\Sigma_1$ with the ``knee" marking the blue-nugget phase.
This universal L-shape evolution track has been confirmed observationally
\citep[][Fig.~7]{barro17}.
%
%\smallskip % obs  
A dissipative ``wet compaction" \citep{db14} is actually required by the 
compactness of the observed massive passive galaxies at $z\!\sim\! 2\!-\!3$ 
\citep[``red nuggets", e.g.][]{dokkum15}, which implies the presence of 
``blue nuggets" as their immediate progenitors.
Indeed, star-forming blue nuggets are observed with consistent masses, 
structure, kinematics and abundance 
\citep[e.g.][]{barro13,dokkum15,barro17}.
A deep-learning study \citep{huertas18}, 
trained on ``observed" mock images of
blue nuggets as identified in the simulations, recognized 
with high confidence blue nuggets in the CANDELS-HST imaging survey of 
$z\!=\!1\!-\!3$ galaxies.

% trigger
The compaction requires dissipative AM loss,  
caused in the simulations either by wet mergers ($\sim\! 40\%$), by colliding 
counter-rotating streams, or other processes. 
% transitions
The compaction marks transitions in the galaxy structural, compositional, 
and kinematic properties, which translate to variations with mass near 
the critical mass for blue nuggets.
For example, the dark-matter dominated center becomes baryon dominated, 
inducing a transition in global shape from a prolate to oblate 
\citep{tomassetti16}.
The growth rate of the central black hole changes from 
slow to fast \citep{dekel19_gold}, inducing a transition from 
supernova feedback to AGN feedback as the main source for quenching.

\smallskip % compaction at a critical mass
Repeated episodes of compactions and quenching attempts explain 
the confinement of SFGs to a narrow Main Sequence \citep{tacchella16_ms}.
\adb{
However, the major compaction events that produce massive bulges
are predicted to occur near a characteristic halo 
mass $\Mv\!\sim\!10^{11.5}\msun$ \citep{zolotov15,tomassetti16}.
This has been confirmed by the deep-learning study of VELA simulations
versus observed CANDELS galaxies \citep{huertas18}, which
detected a preferred stellar mass for the observed blue nuggets
near the golden mass of $\Ms \!\sim\! 10^{9.5-10}\msun$.
}
We argue \citep[][and \se{disc_SN} below]{dekel19_gold} 
that the compaction events are especially pronounced near the 
critical mass because supernova feedback becomes 
inefficient near and above this mass.

%============= % 5.2 
\subsection{Rings stabilized by a massive bulge}
\label{sec:rings}

%-------
%\subsubsection{Discs and rings about a massive bulge}

\Fig{mosaic_v07} shows the evolution of one VELA galaxy 
through the compaction, blue-nugget and post-compaction phases.
The face-on images of projected gas density reflect the 
SFR via the Kennicutt-Schmidt relation.
The blue-nugget at the end of the compaction is the central blob of high 
gas density (second from left).
A highly turbulent rotating disc develops thereafter and is growing in 
extent (third from left), showing a spiral-arm pattern and irregular 
perturbations including giant clumps.
In this violent-disc-instability (VDI) phase the giant clumps and
the gas between them migrate inwards 
\citep[e.g.][]{dsc09}.
Then, the central gas is depleted into star formation and outflows 
\citep{zolotov15} and an extended clumpy ring forms
(fourth panel), continuously fed by incoming cold streams, showing tightly 
wound spiral arms and giant clumps. The ring is sometimes maintained at its 
extended form for several Gyr, with only little inward migration.
The stellar-density maps show how a compact stellar system forms following the
gas compaction process 
\adb{(the stellar nugget seen already in the first image
during the compaction phase reflects earlier minor compactions)}
and how it remains massive and compact as it quenches 
to a passive red nugget. 
These post-compaction discs and rings can be identified with the observed 
massive and extended star-forming rotating and highly turbulent discs showing
giant clumps \citep{genzel08,genzel14_rings,guo15,guo18,forster18a}.

\smallskip % kinemics V not sigma
We learn in
% Fig.~xxx of
\citet{dekel20_ring} that during the compaction
the VELA galaxies evolve from pressure to rotation support, with the median
$\Vrot/\sigma$ growing from near unity to about $4\!\pm\!1$.
It is the rotation velocity that is dramatically growing during the 
compaction, from $\Vrot/\Vv\!=\!0.4\!\pm\!0.1$  
to $\Vrot/\Vv\!= 1.4\!\pm\!0.1$, while 
the velocity dispersion remains at $\sigma/\Vv\!=\!0.4\!\pm\!0.1$.
The kinematic-ratio transition is thus due to an abrupt increase in the gas AM
rather than to a change in turbulence.
This implies that the inflowing high-AM gas is prevented from forming a
long-lived disc in the pre-compaction phase because of an efficient loss
of AM, e.g., due to the merger-driven flips discussed above. 
In turn, the gas seems to retain its incoming high AM in the 
post-compaction phase. 
This motivates our effort to understand the post-compaction survival 
of extended discs and rings by means of AM exchange.

%\smallskip
%We show in 
%Fig.~9xxx of 
%\citet{dekel20_ring} that the transition of VELA galaxies from
%pressure to rotation support is more tightly correlated with the blue-nugget
%phase than with the latest merger or spin-flip event.
%In particular, there are only a small number of cases where the galaxies are 
%non-discs significantly after the major blue-nugget phase.
%This is a possible hint that the formation of a massive central bulge, in the
%mass range where mergers are infrequent, is in most cases sufficient for disc 
%or ring longevity.
%
%While the disciness also shows a general tendency for a transition near the 
%time of a disc flip or a major merger, these correlations involve more
%scatter. Non-discs exist even long after the latest merger,
%indicating that they are not the only mechanisms that disrupt discs.

%%%%%%%%%%%%%%%%%%%%%%%%%%%% 5.3 
%\subsection{A Ring Stabilized by a Central Mass}
%\label{sec:ring_toy}

\smallskip
In \citet{dekel20_ring} we show that the longevity of the rings 
could be understood in the context of a VDI disc if $\delta_{\rm d} \ll 1$,
which could arise from the post-compaction appearance of a massive central 
bulge.
By computing the torques exerted by a tightly wound spiral structure in a thin
disc on its outer parts, we find that the pitch angle can be approximated by 
\be
\tan^2\alpha \sim 4 \delta_{\rm d}^2 
\ee
and the timescale for mass inflow is roughly
\be
t_{\rm inflow,ring} \sim \delta_{\rm d}^{-3}\, \torb\, .
\label{eq:inflow_ring}
\ee
With $\delta_{\rm d}$ small compared to unity, this implies
a very small pitch angle, namely a ring, and a very long inflow time,
much longer than the orbital time.
This is longer than the analogous inflow time of \equ{inflow_disc}, which is
proportional to $\delta_{\rm d}^{-2}$.
The post-compaction massive bulge, that appears above the
critical mass for blue nuggets, reduces $\delta_{\rm d}$, and thus
acts to stabilize the ring against evacuation inwards.\footnote{A similar 
extended long-lived ring would appear about a massive dark-matter
dominated central region, which could be another reason for a reduced
$\delta_{\rm d}$.}

\smallskip
The survival of the ring is determined by the interplay between the inflow
rate of \equ{inflow_ring}, 
the star-formation rate (SFR) within the ring that suppresses the dissipative
inflow \citep{db14},
and the external accretion rate of high-AM gas in spiraling-in streams that
replenish a new outer ring.
If we assume an EdS cosmology (approximately valid at $z>1$),
the ring orbital time is \citep{dekel13} 
$\torb \! \sim\! 0.75 \Gyr\, (1+z)^{-3/2}$.
The accretion timescale is then \citep{dekel13}
$t_{\rm acc}\! \sim\! 40\, \torb\, (1+z)^{-1}$.
In comparison, the inflow rate computed above is
$t_{\rm inflow}\! >\! 40\, t_{\rm orb}$ for $\delta_{\rm d} < 0.3$.
Thus $t_{\rm acc}\! <\! t_{\rm inflow}$ in the ring at all redshifts.  
This implies that the gas accumulates in the ring.

\smallskip % dissipative contraction?
Gas dissipation may speed up the contraction of the ring. The gas turns into
stars on a timescale
$t_{\rm sfr}\! \sim\! \epsilon_{\rm ff}^{-1} t_{\rm ff}
            \! \sim\! 5 \torb$,
where $\epsilon_{\rm ff} \!\sim\!0.01$ is the efficiency of SFR in a 
free-fall time and $t_{\rm ff}$ is the free-fall time in the star-forming 
regions \citep{kdm12}.
We adopted $t_{\rm ff} \sim 0.3\, t_{\rm dyn}$, assuming that stars form in 
clumps that are denser than the mean density of baryons in the ring by a 
factor of $\sim 10$ \citep{ceverino12}.
A similar timescale for SFR is obtained in the VELA simulations.
We learn that $t_{\rm sfr} \ll t_{\rm inflow}$, which implies that the gas in
the ring turns into stars before it has a chance to flow in
such that a dissipative contraction of the ring is not likely.

\begin{figure} % 8
\centering
\includegraphics[width=0.47\textwidth,
trim={{0.09\textwidth} {0.40\textwidth} {0.09\textwidth} {0.31\textwidth}},clip]
{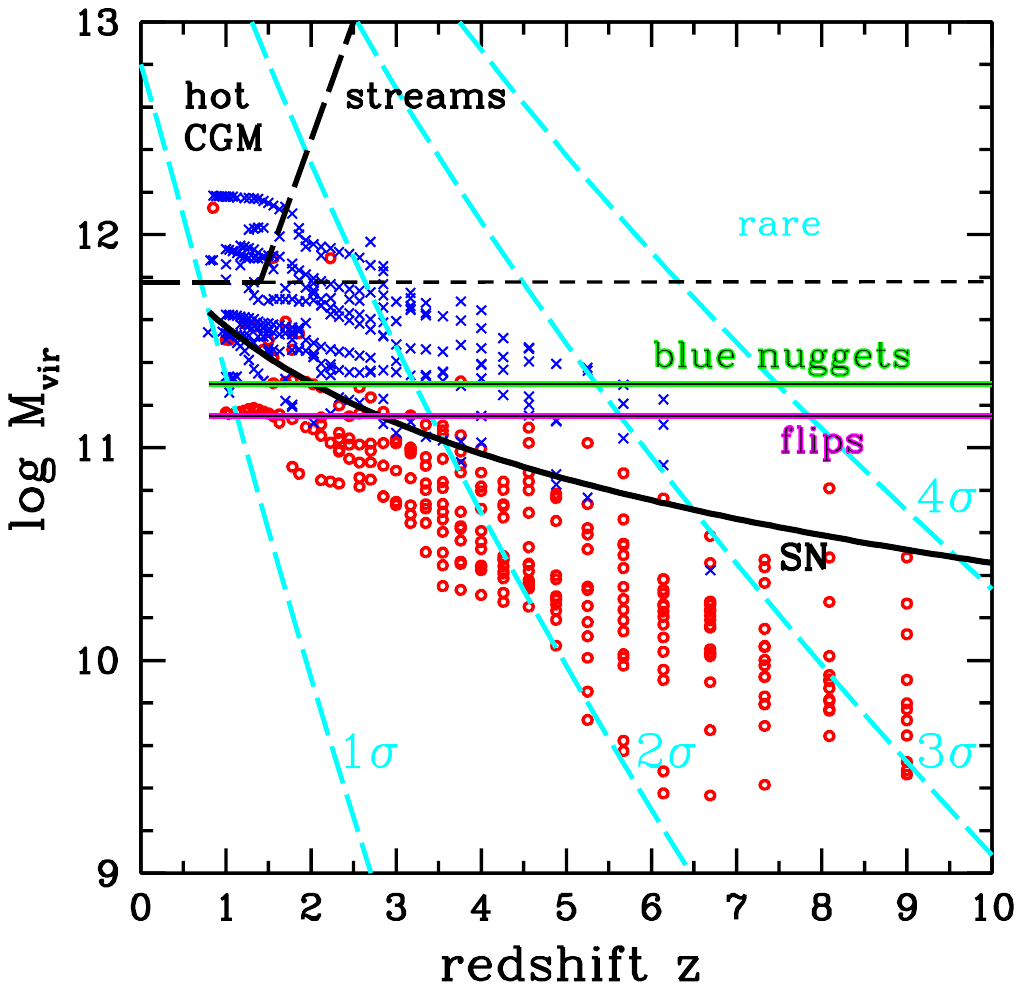}
%\vskip 8.5cm
%\special{psfile="figs/fig8.ps" %disk.ps" 
%         hscale=48 vscale=48 hoffset=-27 voffset=-76}
\caption{
Characteristic masses in the mass-redshift plane.
The magenta line (flip) marks the upper limit for merger-driven disc flipping
in an orbital time, $\Mv \simeq 1.4\times 10^{11}\msun$ of \se{flip_sim} and
\se{flip_toy}.
The green line (BN) refers to the characteristic mass for compaction-driven
blue nugget, $\Mv \simeq 2\times 10^{11}\msun$ \citep{zolotov15}.
The black solid curve (SN) marks the upper limit for effective supernova
feedback, $\Vv \simeq 120\kms$ \citep{ds86}.
The horizontal dashed black line is the characteristic threshold for virial
shock heating and the diagonal dashed black line is the upper limit for
penetrating cold streams \citep{db06}.
The cyan curves refer to the Press-Schechter mass corresponding to $\nu$-sigma
peaks for $\nu=1,2,3,4$; disc flipping is more likely for high-sigma peaks.
Extended discs of star-forming gas are expected to survive above
$\Mv\!\sim\!2\times 10^{11}\msun$ and disrupt at lower masses.
Haloes of $1.4\times 10^{11}\msun$ are expected at a comoving number density
(in $\Mpc^{-3}$) on the order of $10^{-2}$ at $z\!\sim\!1\!-\!3$,
$10^{-3}$ at $z\!\sim\!4\!-\!6$,
and $10^{-4}$ at $z\!\sim\!8$ (\fig{psn}).
The symbols refer to galaxies from the VELA simulations,
selected from snapshots where the gas is either clearly discy
(blue, $\Rd/\Hd\!>\!3.1$ and $\Vrot/\sigma\!>\!3$),
or clearly non-discy
(red, $\Rd/\Hd\!<\!2.8$ and $\Vrot/\sigma\!<\!1.3$), respectively,
each consisting of a quarter of all the snapshots.
Each of the three theoretical curves (blue nuggets, flips and SN)
could serve as a border line between non-discs and discs.
}
\label{fig:Mz}
\end{figure}

%%%%%%%%%%%%%%%%%%%%%%%%%%%%%%%%% 6
\section{On The Role of Supernova feedback and Hot CGM}
\label{sec:disc_SN}

\Fig{Mz} emphasizes the relevant physical characteristic masses in the 
mass-redshift plane. It shows 
(a) the threshold mass for merger-driven disc flips
in an orbital time from \se{flip_sim} and \se{flip_toy}, 
$\Mv \simeq 1.4\times 10^{11}\msun$,
(b) the favored mass for compaction-driven blue nuggets, 
$\Mv \simeq 2\times 10^{11}\msun$ \citep{zolotov15,tomassetti16},
(c) the upper limit for effective supernova feedback, corresponding to
$\Vv \simeq 120\kms$ \citep{ds86},
and (d) the threshold for virial shock heating of the Circum Galactic Medium 
(CGM) and the upper limit
for penetrating cold streams that feed the galaxies \citep{db06,dekel09}.
Also shown are half of the VELA snapshots, selected to be the upper and lower 
quadrants in terms of their disciness as measured by $\Vrot/\sigma$ and
$\Rd/\Hd$. 
One can see that each of the first three lines can serve as a border line
distinguishing between non-disc and disc dominance. 
We argue below that SN feedback has a major role in disrupting discs below the
critical mass in several different ways, direct and indirect, by itself and
through the effects of disc flipping and compactions to blue nuggets.
We then comment on the possible complementary role of a hot CGM in more massive
galaxies.

%=====================  
\subsection{Supernova feedback}
\label{sec:SN-mass}

Simple energetic arguments, using the standard theory for supernova bubbles,
yield 
\adb{a crude}
upper limit for the dark-matter halo mass (actually its virial 
velocity) within which supernova feedback can be effective in heating or 
ejecting the gas and thus suppressing the SFR \citep{ds86}.
The energy deposited in the ISM by supernovae that arise from a 
\adb{burst of stellar mass $\Ms$} is estimated to be 
$E_{\rm SN}\! \sim\! \Ms \Vsn^2$, with $\Vsn\! \sim\! 120 \kms$. 
This is obtained by comparing the cooling and dynamical timescales for the
adiabatic phase of SN bubbles and for the star-formation rate, respectively.
For the supernova energy to heat or eject most of the gas of mass $\Mg$
that has accreted into the galaxy it should be comparable to the binding
energy of this gas in the dark-matter halo potential well,
$E_{\rm CGM}\! \sim\! \Mg \Vv^2$, where $\Vv^2\! =\! {G\Mv}/{\Rv}$
characterizes the halo potential well.
At the peak of star-formation efficiency, if a large fraction of the 
cumulative accreted gas turned into stars with little ejection, one can
approximate $\Ms\!\sim\!\Mg$. Then, 
comparing $E_{\rm SN}$ and $E_{\rm CGM}$, 
one obtains an upper limit for effective SN feedback at 
$\Vv \!\sim\! \Vsn \!\sim\! 120 \kms$.
The corresponding halo mass in the EdS regime ($z\!>\!1$) is
\citep[e.g.][Appendix A2]{db06}
\be
\Mv \simeq 6 \times 10^{11} \msun V_{{\rm v},100}^3\,(1+z)^{-3/2} \, ,
\ee
where $\Vv=100\kms\,V_{{\rm v},100}$.
This SN critical mass is plotted in \fig{Mz}.
Despite the redshift dependence of the mass,
it is in the ballpark of the mass threshold for transition from non-discs
to discs (\se{flip_sim}) at all redshifts, 
which is also similar to the critical mass for blue nuggets (\se{compaction}).
%We summarize next the possible ways by which SN feedback below the critical
%mass may be responsible for the transition from non-discs to discs, 
%in a direct or an indirect way.

\smallskip % slope in the SN zone - and flips
We saw in \se{flip_toy} that the characteristic mass for disc flips arises 
from the relation between stellar and halo mass in the SN zone, below the 
critical mass, \equ{behroozi19}. 
The above simple energetics argument for SN feedback can also be used to 
explain the slope of this relation.
When $\Vv \!<\! \Vsn$, the shallow potential well allows
significant gas ejection such that $\Ms \!\ll\! \Mg$.
If the gas mass that has been accreted into the halo is 
roughly proportional to the halo mass, $\Mg \!\prop\! \Mv$,
by comparing $E_{\rm SN}$ and $E_{\rm CGM}$ one obtains \citep{dw03}
\be
\frac{\Ms}{\Mv} \prop \Vv^2 \prop \Mv^{2/3} \, .
\ee
This mass dependence is in the ballpark of the relation deduced
from observations via abundance matching with LCDM haloes
\citep{moster10,behroozi13},
while the analysis of \citet{behroozi19} as summarized in
\equ{behroozi19} indicates a slightly steeper slope, closer to unity.
Thus, SN feedback seems to have an indirect but important role in determining 
the critical mass for disc flips (\se{flip_sim} and \se{flip_toy}).

\smallskip % SN feedback boosted in mergers. help disrupting gas discs.
Each of the major mergers that cause spin flips below the critical mass
also generates a starburst, which leads to a boosted SN feedback.  
This SN feedback may drive excessive turbulence in the disc, which puffs it up 
and possibly disrupt it.
% winds halting infall of hi AM
Furthermore, the resultant supernova-driven winds suppress the inflow into 
the galaxy, as has been demonstrated by \citet[][Fig. 5]{tollet19}. 
Since the inflowing gas is typically of a higher AM than the existing galaxy, 
the SN-driven winds limit the growth of an extended disc.

\smallskip % fountain
Another possible mechanism for losing AM involves an ``inverse fountain", 
where  supernovae at the outskirts of the disc preferentially remove 
high-AM gas 
(that may return after losing AM in the inner CGM), thus trimming the galaxy
from its extended gas disc (Degraf et al., in prep.). 
The fountain becomes of an opposite, more standard effect on the
disc viability after the compaction to a blue nugget, where the SFR becomes 
concentrated in the central regions such that SN feedback preferentially
removes low-AM gas, which increases the net specific AM in the disc.   

\smallskip % compaction at a critical mass
Finally, it has been argued that SN feedback is responsible for confining the
major compaction events to the golden halo mass of a few $\times 10^{11}\msun$
\citep{dekel19_gold}.
Compaction events do occur also at smaller masses, where they help confining 
the star-forming galaxies into a narrow main sequence \citep{tacchella16_ms}.  
However, SN winds in the heart of the SN zone prevent these compactions to
proceed deep into the galaxy central kiloparsec. 
They do not cause significant long-term post-compaction quenching as the gas 
is rapidly replenished by and the centres remain dark-matter dominated. 
Only near the upper-limit critical mass for SN feedback the winds become 
ineffective and the compaction process can proceed deep into the galaxy 
centre, generating a blue nugget that triggers a lasting central quenching.
Given that the blue nuggets seem to be a major driver of an extended disc
or ring (\se{compaction_ring}, \se{rings}), 
we learn that the SN feedback has an indirect but important role in 
determining the mass threshold for the formation of long-lived rings.

%--------------
\subsection{Hot CGM}

% hot cgm
Another characteristic mass of a complementary role here 
is the upper limit for the halo 
mass within which cold inflow can supply gas for an extended
star-forming disc or ring
\citep[originally][]{ro77,silk77,binney77}.
It is obtained by comparing the gas radiative cooling time to the 
dynamical time for gas inflow.
The key is whether the shock that forms at the halo virial radius,
behind which the gas heats to the virial temperature,
can be supported against gravitational collapse.
For this, its cooling time has to be longer than the
dynamical time for gas compression behind the shock.
A shock-stability analysis reveals a threshold mass for a hot CGM
\citep{bd03,db06}, marked in \fig{Mz},
on the order of
\be
\Mv \sim 5\times 10^{11}\msun \, ,
\ee
roughly independent of redshift. 
The associated uncertainty and scatter between galaxies are rather large,
spanning about a decade, e.g., reflecting uncertainties in metallicity
and the location within the halo where the shock stability is evaluated.

\smallskip % streams
At $z\! \geq\! 2$, narrow cold streams are expected to
penetrate through the otherwise hot CGM and supply gas for discs or rings
even in haloes above the critical mass for shock heating
\citep[][Figure 7]{db06}.
These predictions have been confirmed in cosmological
simulations \citep{keres05,ocvirk08,nelson13,nelson16},
and observationally based on abundance matching of galaxies to dark-matter 
haloes in a LCDM cosmology \citep{behroozi19}.
For shutdown of gas supply that would starve the discs or rings,
the CGM has to be kept hot. This
can be caused above the upper limit for cold streams
by gravitational heating due to accreting mass 
\citep{bdn07,db08,khochfar08},
as well as by AGN feedback 
\citep[e.g.][]{croton06,cattaneo07,ciotti07,dubois11},
as summarized in \citet{dekel19_gold}.

\smallskip % quenching above the CGM threshold, and morphology by dry mergers
At low redshifts, the hot CGM in haloes of high masses has a role in quenching 
gas discs and star formation
but it is not supposed to be a direct cause for morphological changes by
disc disruption, thus allowing quenched stellar discs.
However, while the disc gas is consumed into stars and is possibly ejected by
winds, the hot CGM suppresses the replenishment of the disc by fresh gas with
high AM and thus suppresses the long-term existence of very 
extended gaseous star-forming discs in massive galaxies.
Above the threshold mass for hot CGM one expects quenched stellar systems,
which gradually become less discy under dry major and minor mergers.
On the other hand, at $z\!>\!2$, the penetrating cold streams allow 
continuous gas supply for an extended disc or ring also in haloes above 
the critical mass for shock heating, where according to our current findings
they are not frequently disrupted by spin flips.

%%%%%%%%%%%%%%%%%%%%%%%% 7
\section{Other simulations \& observations}
\label{sec:disc_other}

Here we compare our results to related simulation work and tentatively
address existing observational results, looking forward to future observations,
e.g., by JWST, at the relevant low masses $\Ms \leq 10^9\msun$ and at high 
redshifts.

%================ 7.1
\subsection{Other Simulations}
\label{sec:other_sims}

As a relevant background,
spin flips of dark-matter haloes were studied in cosmological simulations
by \citet{bett12} and \citet{bett16}, who found that they could indeed be 
associated with either major mergers or with non-major mergers and other 
changes in the AM feeding pattern and rate. 
First steps in addressing the effects of mergers of cosmic-web filaments,
using excersion-set theory for saddle-point mergers,
were made by \citet{musso18}.
Also related are the correlations of galactic spins with the
filaments and walls of the cosmic web as analyzed in cosmological simulations
by \citet{kraljic20}.
The spins of the dark-matter haloes may provide clues for what one should 
expect for the disc spins, but one should be aware of the significant AM gains
and losses of the instreaming baryons as they penetrate from outside the halo 
to the inner galaxy \citep{danovich15}, which tends to smear out most of the 
one-to-one correlation between the disc and halo spins \citep{jiang19_spin}. 

\smallskip %El-Badry+18
\citet{elbadry18a} studied the viability of discs as a function of stellar mass
at $z=0$ in 24 FIRE zoom-in cosmological simulations, 
spanning stellar masses of $\Ms=10^{6-11}\msun$.
They find at $z\!=\!0$ a trend with mass similar to our finding at 
$z\!\geq\! 0.8$,
with a transition from dominance by non-discs to discs near a similar critical 
mass, $\Ms \sim 10^9\msun$.
They also realize that stellar feedback has a major role in reducing the
rotational support in low-mass galaxies, owing to the suppression of the
accretion of high-AM gas and to driving large non-circular gas motions.
We argued that, on top of these direct effects of SN feedback, it affects the
transition from non-discs to discs in two other major ways, namely trough  
frequent merger-driven disc flips in low-mass
galaxies, and through the compactions to massive bulges above the same critical
mass, which make the extended rings long-lived.

\begin{figure} % 9
\centering
\includegraphics[width=0.47\textwidth,
trim={{0.09\textwidth} {0.40\textwidth} {0.09\textwidth} {0.31\textwidth}},clip]
{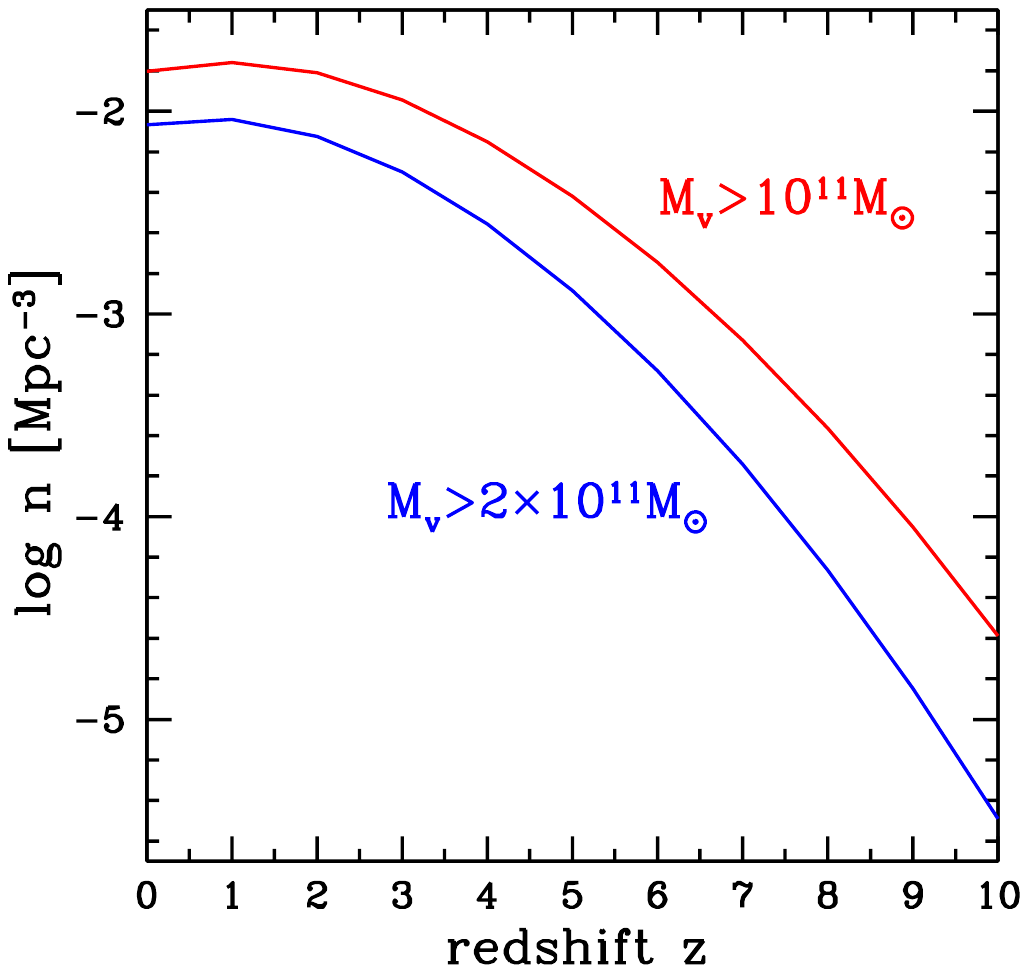}
%\vskip 7.8cm
%\special{psfile="figs/fig9.ps" %psn.ps" 
%        hscale=44 vscale=44 hoffset=-27 voffset=-76}
\caption{
Comoving number density for discs as a function of redshift, as derived
by the Press-Schechter formalism for haloes above the threshold mass.
Shown are curves for $\Mv\!>\!10^{11}\msun$ and $>\!2\times 10^{11}\msun$, as
could be deduced from \fig{disc_Mz_bins}.
}
\label{fig:psn}
\end{figure}

\smallskip
\citet{elbadry18a} 
also point out that the stellar systems tend to be even less rotationally
supported than the gas, particularly at low masses. This is consistent with
or findings in \citet{danovich15}.
Indeed, 
\citet{wheeler19} performed very high resolution simulations of dwarf galaxies,
and found stellar kinematics with very low rotational support, 
$\Vrot/\sigma \leq 0.5$, for a few dwarfs in the range
$\Ms = 10^{5-7} \msun$. 
%This is for the stellar kinematics, 
%which is expected to be of a lower rotational support.
It will be useful to obtain analogous kinematics for the gas component.

%================ 7.2
\subsection{Tentative comparison to observations}
\label{sec:obs}

The existing observations of gas kinematics are largely limited either to low 
redshifts or to masses above the predicted threshold for disc dominance. 
Furthermore, the selection is typically biased toward disc morphologies, 
which makes it hard to explore the predicted non-disc zone below 
$\Ms\! \sim\! 10^9\msun$. 
We summarize here preliminary comparisons to existing observations.
More suitable observations will probably have to wait for studies using future 
tools such as JWST. 

\begin{figure*} % 10
\centering
\includegraphics[width=0.9\textwidth]{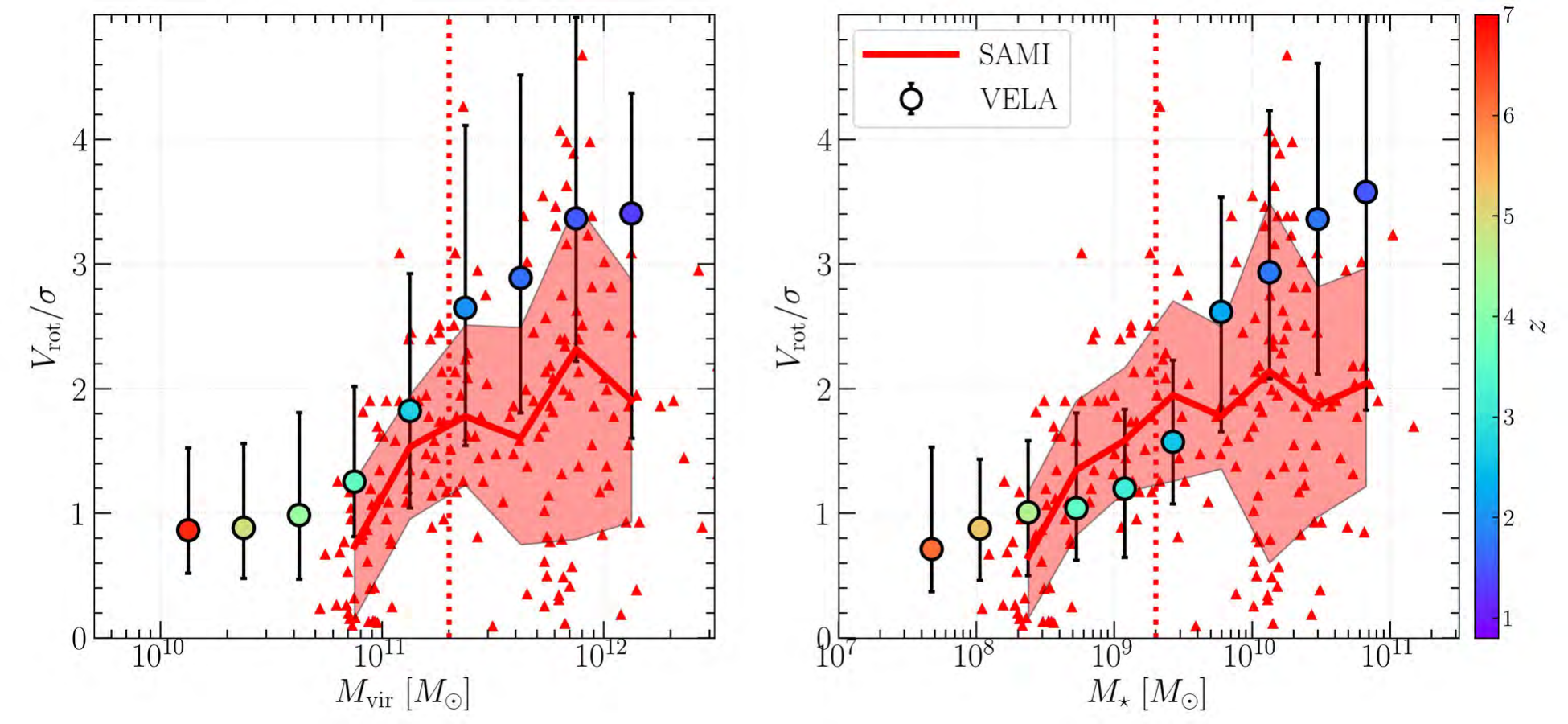}
%\vskip 7.4cm
%\special{psfile="figs/fig10.eps" %sami_VrotMstar.eps" 
%          hscale=48 vscale=48 hoffset=20 voffset=0}
\caption{
Preliminary comparison to observations.
Disciness as measured by $\Vrot/\sigma$ as a function of virial mass (left)
and stellar mass (right).
Shown for the VELA simulations are the medians (circles) and $1\sigma$
error bars (16 and 84 percentiles).
The observational estimates from the SAMI survey \citep{cortese14}
are shown as red triangles with the curve and shaded areas marking
the median and $1\sigma$ scatter.
While the observations are at low $z$, the simulations span a range of
redshifts (color).
The systematic increase of disciness with mass is similar in the simulations
and the observations, both showing a sharp increase near
$\Mv \sim 2\times 10^{11}\msun$.
The simulations slightly overestimate the observations at large masses
but only at the level of $1\sigma$ or less.
It may be due to the different ways $\Vrot$ and $\sigma$ are evaluated
with limited resolution in 2D (in the observations)
versus 3D (in the simulations).
}
\label{fig:sami}
\end{figure*}

%---------------- 
\subsubsection{Predicted abundance of disc galaxies}

Given the threshold mass for discs,
the expected abundance of disc galaxies in a given redshift
can be estimated by the number density of haloes above the threshold mass.
\Fig{psn} shows the comoving number density of haloes of 
either $\Mv\!>\!2\times 10^{11}\msun$ or $\Mv\!>\!10^{11}\msun$
as a function of redshift based on the Press-Schechter formalism 
\citep{press74}, assuming a standard LCDM cosmology 
(with $\omm\!=\!0.3$, $\oml\!=\!0.7$, $\omb\!=\!0.045$, 
$h\!=\!0.7$ and $\sigma_8\!=\!0.82$).
The Sheth-Tormen version of the formalism is used \citep{sheth02}, 
as summarized in the appendix of \citet{db06}.
The functional fit of \citep{bbks86} to the density fluctuation power spectrum 
is used, with the correction for baryons proposed by \citet{sugiyama95}.
This approximation is good to better than 5\% compared to the more elaborate
computation of \citet{eisenstein99}.
The mass threshold implies a comoving number density of
discs $n\!>\!10^{-2}\Mpc^{-3}$ in the redshift range $z\!=\!0\!-\!2$,
and $n\!>\! 2.8\times 10^{-3},\, 5.2\times 10^{-4},\, 
3.2\times 10^{-5}\Mpc^{-3}$ at $z\!\simeq\!4,6,10$ respectively.
The black curves in \fig{disc_Mz_bins} refer to the Press-Schechter
$\nu\sigma$ peaks, for $\nu=1,2,3,4$ from left to right respectively.

\subsubsection{Gas kinematics at low redshift: SAMI}

% SAMI
The SAMI survey provides HI gas kinematics for galaxies at low redshift
in the mass range $\Ms\!=\!10^{8-11}\msun$ \citep[][Fig. 2]{cortese14}. 
This survey was supposedly selected blind to morphology, though in practice
it seems to be biased at low masses towards late type discs, Sc or later
(their Fig. 2, bottom panels).
\Fig{sami} shows the disciness, as measured by $\Vrot/\sigma$,
versus virial mass and versus stellar mass, for the VELA simulations 
compared to the observational estimates from the SAMI survey. 
%\citep[][Fig. 2]{cortese14}.
The data provided are $\Vrot$ and $\sigma$ as a function of stellar mass.
The virial halo mass was determined from the stellar mass using our functional
fit to the $\Ms\!-\!\Mv$ relation by \citet{rodriguez17}, 
$\Mv\! =\! 25\,\Ms\, ( {\Ms}_{10.6}^{-0.5} \!+\! {\Ms}_{10.6}^{1.5} )$,
where ${\Ms}_{10.6}\!=\!\Ms/10^{10.6}\msun$. 
The stellar-to-halo mass ratio is slightly higher in the VELA3 simulations,
which introduces a horizontal uncertainty in the comparison
between the simulations and observations.
We note that the observations are at low $z$, while the simulations span a 
range of redshifts, generally monotonic with mass as the galaxies grow in
time, such that the low-mass tail largely refers to $z\!>\!4$.

\smallskip
We see a qualitative agreement between theory and observations, both showing
a similar systematic increase of disciness with mass, with a steep drop
below $\Mv\! \sim\! 2\times 10^{11}\msun$.
The simulation disciness seems to be marginally higher than the observational 
estimates at large masses, though this difference is only at the level of 
$1\sigma$ or less.
This may well be due to the different ways $\Vrot$ and 
$\sigma$ are evaluated in 2D (in the observations) versus 3D 
(in the simulations). In particular, because of the limited resolution, the
observations naturally tend to underestimate $\Vrot$ and overestimate $\sigma$.
The observational analyses attempt to correct for this effect, but this may be
an under-correction or an over-correction, leaving a significant uncertainty in
the measured $\Vrot/\sigma$.

\smallskip
\citet[][Fig. 15]{elbadry18a} show the observed Tully-Fisher relation
based on the gas rotation velocities from the SAMI survey.
Here, too, the data show a steeper slope in the range 
$\Ms\! \sim\! 10^{8-9}\msun$,
crudely $\Vrot\! \prop\! \Ms^{0.7}$ or steeper,
compared to a log slope $\sim\! 0.3$ in the higher-mass range 
$\sim\! 10^{9-10}\msun$, just above the critical mass.
These are to be compared to the much flatter slope that one would have
obtained by assuming rotational support, $\Vrot \prop \Vv$,
and $\Ms \prop \Mv^2$ from \equ{behroozi19}, which would give 
$\Vrot \prop \Mv^{1/3} \prop \Ms^{1/6}$.
The steeper observed slope is qualitatively consistent with our predicted lower
rotational support at lower masses, particularly below the critical mass.

\subsubsection{Gas kinematics above the critical mass: Simons~et~al.}

\begin{figure} %11
\centering
\includegraphics[width=0.47\textwidth,
trim={{0.09\textwidth} {0.34\textwidth} {0.09\textwidth} {0.25\textwidth}},clip]
{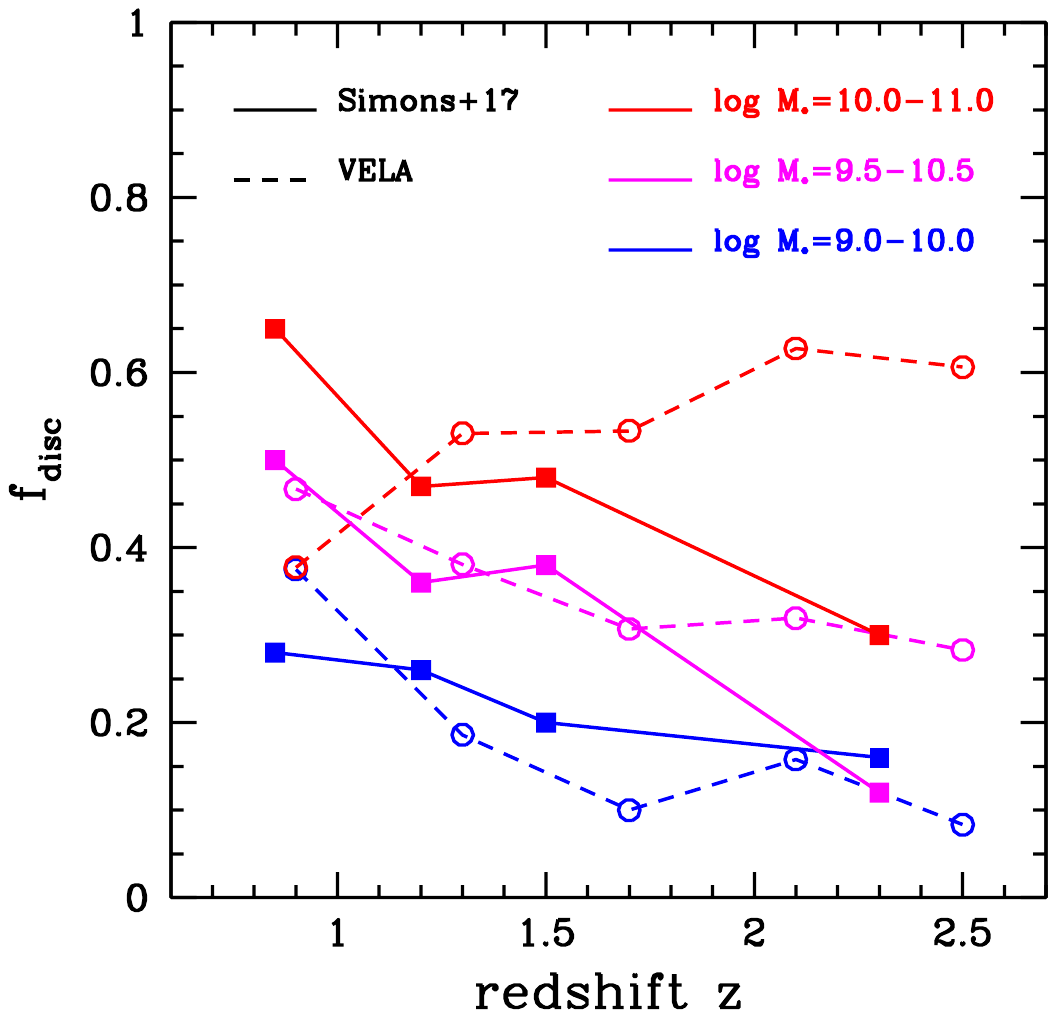}
%\vskip 8.5cm
%\special{psfile="figs/fig11.ps" %simons_short.ps" 
%        hscale=48 vscale=48 hoffset=-27 voffset=-76}
\caption{
Preliminary comparison to observations.
Fraction of discs ($\Vrot/\sigma>3$) as a function of redshift
in the VELA simulations (open circles, dashed lines) compared to the
observational determinations by \citet[][Fig. 4b]{simons17} (filled squares,
solid lines),
for three stellar mass bins, in the overlapping redshift range.
There is a good agreement for $\Ms = 10^{9.0-10.5}$, showing a slow rise with
time of the disc fraction in this redshift range.
}
\label{fig:simons}
\end{figure}

% Simons M
\citet[][Fig. 11]{kassin12} and \citet[][Fig. 4]{simons17} show their
estimate of the fraction of observed disc galaxies with $\Vrot/\sigma\!>\!3$ as
a function of stellar mass and redshift.
The former reference spans the range $\Ms\!=\!10^{8-10.7}\msun$ at 
$z\!=\!0.2\!-\!1.1$ and
the latter more comprehensive study covers $\Ms\!=\!10^{9-11}\msun$ at 
$z\!=\!0.2\!-\!2.4$.
\Fig{simons} shows the fraction of discs ($\Vrot/\sigma\!>\!3$) as a function of
redshift in stellar mass bins,
in the VELA simulations compared to the observational estimates by
\citet[][Fig. 4b]{simons17}.
The data are for three stellar mass bins, in the limited overlapping redshift
range.
The observations show a systematic mass dependence at any given redshift,
crudely $f_{\rm disc}\! \prop\! \Ms^{0.3-0.4}$.
This is similar to what we find in the VELA
simulations in the same mass and redshift range.
Note that the mass range studied by \citet{simons17} is limited to near and
above  the predicted transition mass.

\smallskip % Simons z
The observations by \citet{simons17} also indicate in the given redshift range
a redshift dependence of $f_{\rm disc}\! \prop\! (1+z)^{0.8}$
for $\Ms\!=\!10^{9-10}\msun$, which seems to be in apparent conflict with our
general prediction of no significant redshift dependence over the larger
redshift range $z\!=\!1\!-\!8$.
However, when focusing on the limited redshift range $z\!=\!0.8\!-\!2.4$, where
there are both simulated and observed galaxies, as in \fig{simons},
the simulations predict a similar redshift dependence for the two low mass
bins.
In the most massive bin the simulations predict a higher disc fraction than
estimated observationally at $z\! \sim\! 2$.
On one hand, this difference may reflect a flaw in the simulations.
For example, as mentioned in \se{flip_toy},
a redshift dependence of the stellar-to-halo mass ratio, e.g., as deduced by
\citet{rodriguez17} contrary to \citet{behroozi19}, would have introduced a
redshift dependence in the disc fraction along the lines of the observations
over a large redshift range.
On the other hand, the difference in the high-mass bin at $z\!\sim\! 2$ could
reflect differences in the ways the measurements of $\Vrot$ and $\sigma$ are
performed in the simulations and observations, and especially the effects of
resolution on the kinematic ratio derived from observations.

\subsubsection{Gas kinematics in low redshift discs: THINGS}

\citet{elbadry18a} summarize observations of HI gas specific AM.
In their Fig.~13, they show the specific AM of HI gas, $j_{\rm HI}$,
versus baryonic mass, as measured in the HI surveys THINGS \citep{obreschkow14} 
and LITTLE THINGS \citep{butler17} as well as from \citet{bosch01} and
\citet{chowdhury17}. 
These data indicate a marginal bend of the $j_{\rm HI}\!-\!M_{\rm bar}$ 
relation  near $M_{\rm bar}\! \sim\! 10^9\msun$, as predicted. 
The logarithmic slope of the relation above 
this mass is $\simeq\! 0.4$ and below this mass it is $\simeq\! 0.75$.
These slopes can be compared to what one would have deduced from the
halo virial relations.
Assuming that the effective gas radius and rotation velocity are
constant fractions of $\Rv$ and $\Vv$, one obtains 
$j_{\rm gas}\! \simeq\! \Vv \Rv\! \prop\! \Mv^{2/3}$.
Adopting $M_{\rm bar}\! \prop\! \Mv^{1.7}$ from \equ{MdMv},
the expected relation becomes
$j_{\rm gas}\! \prop\! M_{\rm bar}^{0.4}$.
This is consistent with the observed slope just above the bend, 
while the slope below the bend is steeper, indicating less rotational 
support for lower masses.
When plotted against $\Mv$, the change of slope is expected to become more 
pronounced.
This is qualitatively along the lines of our predictions of a gradient of 
rotational support as a function of mass, with the transition near 
$\Ms\! \lsim\! 10^9\msun$.
Unfortunately, the observed galaxies were selected to have a disc morphology,
so the sample misses the low-rotation non-discs at low masses, and may thus
miss much of the predicted effect.

\subsubsection{Stellar kinematics of dwarfs: Wheeler et al.}

\citet{wheeler17} performed a Bayesian analysis of the observed $\Vrot/\sigma$ 
for stars in 40 dwarf galaxies in the local volume over 
$\Ms\!=\!10^{3.5-8}\msun$, 
namely below the predicted critical mass for rotation-supported discs.
Their sample includes both dwarf spheroidals and star-forming dwarf irregulars.
They find that 80 percent of the stellar systems in these dwarfs are 
dispersion supported. 
They see no clear trend between the stellar $\Vrot/\sigma$ and $\Ms$ in the 
mass range $\Ms\!=\!10^{3.5-6}\msun$, which we note is well below the threshold 
mass for disc formation.
They also found that in four FIRE simulated isolated galaxies in the range
$\Mv\!=\!10^{9-10}\msun$ the values of the stellar $\Vrot/\sigma$ are 
consistent with the observed values.
While this is a hint,
these results are not straightforwardly comparable to our predictions
as extending our analysis of gaseous discs to stellar kinematics is beyond the
scope of our current paper.

\subsubsection{Rings in massive galaxies at high $z$: Genzel et al.}

We predict that above the threshold mass the rotating gas discs are in many
cases clumpy rings surrounding a massive body.
When observing such rings, one should be careful to correct for dust, which may
introduce false rings.
More details concerning the expected ring properties are provided in
\citet{dekel20_ring}.

\smallskip
At $z\!\sim\! 2$, \citet{genzel14_rings} detected a significant 
fraction of massive galaxies with extended $H_\alpha$ star-forming rings, 
most of which surrounding a central massive stellar bulge.
Genzel et al. (2019, in prep.), following \citet{genzel17}, 
who determined rotation curves for 40 massive star-forming galaxies at
$z\!=\!0.6\!-\!2.7$ using data from the 3D-HST/KMOS$^{\rm 3D}$/SINFONI/NOEMA 
surveys, verify the common existence of extended gas rings. 
Some of the rings surround massive compact bulges, typically with little
dark-matter mass within the effective radius,
while other rings surround less massive bulges but with higher central
dark-matter masses.  
In both cases this is consistent with the presence of a large central mass.

\smallskip
In an ongoing work (Zhiyuan, Giavalisco et al. 2020, in prep.), 
we find many star-forming ring galaxies in medium-deep HST images 
observed in the U-F366W bandpass in the GOODS-S field, 
complementing them with the 
corresponding B-F453W and H-F160W images, within the CANDELS survey.
Our preliminary inspections indicate that, among the galaxies of 
$\Ms\!>\!10^{9.5}\msun$ at $z\!=\!0.5-2$, a significant fraction show 
star-forming, clumpy rings, typically surrounding a massive bulge, 
which is either star forming or quenched.

\smallskip
At low redshifts, \citet{salim12} found that most ``Green-Valley" 
galaxies, at the early stages of their quenching process,  
consist of massive quenched bulges surrounded by star-forming rings.
These rings may be related to the high-$z$ rings discussed above as
the Green Valley may be the low-$z$ analog of the post-compaction phase
that triggers the quenching at high $z$ 
\citep{tacchella16_ms}.

%%%%%%%%%%%%%%%%%%%%% 8
\section{Conclusion}
\label{sec:conc}

We predict
that galactic gas discs are likely to survive only in dark-matter haloes of 
mass above a threshold of $\sim\!2 \times 10^{11}\msun$, corresponding to a 
stellar mass of $\sim\!10^9\msun$, with only little dependence on
redshift.
This is derived by analytic toy models and confirmed in zoom-in hydrodynamical
cosmological simulations.
In haloes of lower masses, the gas does not tend to settle into an extended
long-lived rotating disc, as several different mechanisms act to reduce the 
angular momentum and thus disrupt the disc.

\smallskip
Most importantly, 
the angular momentum is predicted to flip on a timescale shorter than 
the orbital timescale due to mergers associated with a pattern-change in the
cosmic-web streams that feed the galaxy with angular momentum.
We show analytically why the mass threshold for flips is due to the 
dependence of the baryon-to-halo mass ratio on mass. 
The redshift dependencies of the ratio of accretion-to-orbital 
timescales and of the stellar fraction in the galaxy balance each other,
giving rise to a mass threshold that hardly varies with redshift.

\smallskip
Secondly, in this pre-compaction low-mass regime,
violent disc instability exerts torques that drive angular-momentum out and
mass in, thus making the disc shrink in a few orbital times.
Furthermore, in this regime the central dark-matter and stellar system tend 
to be prolate and thus capable of producing torques that reduce the angular 
momentum of the incoming new gas.

\smallskip
We argue that supernova feedback has an important role in the disruption of
discs below the critical mass. 
It determines the stellar-to-halo mass ratio that affects the merger rate, 
it stirs up turbulence that puffs up the disc and it suppresses the supply 
of new gas with high angular momentum, possibly even ejecting high-AM gas 
from the disc outskirts.
Supernova feedback also confines the major compaction events to near
and above the critical mass and thus stabilize the extended rings above 
this threshold.

\smallskip
Our simulations indicate only a weak redshift dependence of disc survival.
This is associated with an insensitivity of the stellar-to-halo mass ratio to
redshift, as implied by certain abundance matching analyses \citep{behroozi19}.
A preferred occurance of discs of a given mass at late times could arise
instead if the stellar-to-halo mass ratio is actually increasing with time
\citep{rodriguez17}.
The weak redshift dependence is also
a manifestation of the weak redshift dependence of the effectiveness of
supernova feedback, which is primarily determined by the depth of the potential
well, a function of the halo virial velocity which is primarily a function
of halo mass.

\smallskip
Above the critical mass, the disruptive mergers are less frequent and are not
necessarily associated with a change in the pattern of the feeding streams, 
allowing the discs to survive for several orbital times.
In parallel, the effects of supernova feedback are reduced due to the depth of
the halo potential well.

\smallskip
Potential disc disruption by inward mass transport due to violent disc 
instability is limited to below a similar critical mass.
In the post-compaction regime, when the galaxy mass is typically above the 
threshold mass, the simulations show that the inflowing high-AM streams settle 
into long-lived extended discs that evolve into rings.
By computing the torques exerted by a tightly wound spiral structure on
an extended ring, we show in \citet{dekel20_ring} that the timescale for 
mass inflow is roughly $\delta_{\rm d}^{-3}\torb$, 
where $\delta_{\rm d}$ is the disc-to-total mass ratio.
This implies that the post-compaction massive bulge, that appears above the
critical mass for blue nuggets, acts to stabilize the ring against shrinkage by
reducing $\delta_{\rm d}$ to values well below unity.
%A similar extended long-lived ring would appear about a massive dark-matter 
%dominated central region, which could be another reason for a reduced 
%$\delta_{\rm d}$.
Preliminary observational studies in progress indeed indicate detections of 
gaseous star-forming rings around massive bulges or dark-matter dominated 
centers in a large fraction of $z\! \sim\! 1\!-\!2$ galaxies above the 
threshold mass, consistent with the predicted role of compaction in the
longevity of extended rings. 

\smallskip
The threshold of $\Mv\!\sim\!1.4\!\times\!10^{11}\msun$ 
implies a disc comoving number density of 
$n\!\sim\! 10^{-2}, 10^{-3}, 10^{-5}\Mpc^{-3}$ at $z\!=\!2,6,10$. 
The results from existing kinematical observations are consistent with the
predicted mass dependence and threshold mass for long-lived gas discs.
These include the studies based on the SAMI survey \citep{cortese14}
and the analysis of observations by \citet{simons17}.
However, the current observations hardly explore the mass range below 
$\Ms\! \sim\! 10^9\msun$ at moderate and high redshifts, where our simulations
and analysis indicate that disc disruption is expected
to dominate. In several cases (e.g. THINGS) the selection is biased toward
discs, thus disabling a study of non-discs at low masses.
Future observations, e.g. with JWST, may allow an exploration of
this regime.

%\be
%\begin{aligned}
%\Rs =& (4\pi/3)^{-1/12} G^{-1/4} \rho_0^{-7/12}
%       s^{1/2} \eta^{-1/2} \Delta^{-1/2} \tilde\Delta^{-1/12} \\ 
%     & \times \Ma^{-1/2} \Mv^{1/3} a^{1/2} \delta^{-1} .
%\end{aligned}
%\ee
%\be
%\boxed{
%\frac{\Rs}{\Rv} = 0.10\, s_{0.015}^{1/2} \eta_{0.4}^{-1/2} 
%            \Delta_{200}^{-1/2} \tilde\Delta_{200}^{1/4} \Ma_{1.5}^{-1/2}
%            \delta_{75}^{-1/2} a_{1/3}^{-1/2} .
%}
%\label{eq:Rs-delta}
%\ee

%%%%%%%%%%%%%
\section*{Acknowledgments}

We are grateful for stimulating discussions with Sandy Faber,
Reinhard Genzel, David Koo and Doug Lin. 
This work was partly supported by the grants 
Germany-Israel GIF I-1341-303.7/2016, Germany-Israel DIP STE1869/2-1
GE625/17-1, I-CORE Program of the PBC/ISF 1829/12, ISF 857/14,
US-Israel BSF 2014-273, and NSF AST-1405962.
The cosmological VELA simulations were performed at the National Energy
Research Scientific Computing Center (NERSC) at Lawrence Berkeley National
Laboratory, and at NASA Advanced Supercomputing (NAS) at NASA Ames Research
Center. Development and analysis have been performed in the astro cluster at
HU.

%%%%%%%%%%%%%%%%%%%%%%%%%%%%%%%%%%%%%%%%%%%%%
\bibliographystyle{mn2e}
\bibliography{flip}

\begin{thebibliography}{104}
\expandafter\ifx\csname natexlab\endcsname\relax\def\natexlab#1{#1}\fi

\bibitem[{{Agertz} {et~al}\mbox{.}(2013){Agertz}, {Kravtsov}, {Leitner}, \&
  {Gnedin}}]{agertz13}
{Agertz} O., {Kravtsov} A.~V., {Leitner} S.~N., {Gnedin} N.~Y., 2013, \apj,
  770, 25

\bibitem[{{Bardeen} {et~al}\mbox{.}(1986){Bardeen}, {Bond}, {Kaiser}, \&
  {Szalay}}]{bbks86}
{Bardeen} J.~M., {Bond} J.~R., {Kaiser} N., {Szalay} A.~S., 1986, \apj, 304, 15

\bibitem[{{Barro} {et~al}\mbox{.}(2017){Barro}, {Faber}, {Koo}, {Dekel},
  {Fang}, {Trump}, {P{\'e}rez-Gonz{\'a}lez}, {Pacifici}, {Primack},
  {Somerville}, {Yan}, {Guo}, {Liu}, {Ceverino}, {Kocevski}, \&
  {McGrath}}]{barro17}
{Barro} G. {et~al.}, 2017, \apj, 840, 47

\bibitem[{{Barro} {et~al}\mbox{.}(2013){Barro}, {Faber}, {Perez-Gonzalez},
  {Koo}, {Williams}, {Kocevski}, {...}, {Dekel}, \& {et al.,}}]{barro13}
{Barro} G. {et~al.}, 2013, \apj, 765, 104

\bibitem[{{Behroozi} {et~al}\mbox{.}(2019){Behroozi}, {Wechsler}, {Hearin}, \&
  {Conroy}}]{behroozi19}
{Behroozi} P., {Wechsler} R.~H., {Hearin} A.~P., {Conroy} C., 2019, \mnras,
  488, 3143

\bibitem[{{Behroozi}, {Wechsler} \& {Conroy}(2013){Behroozi}, {Wechsler}, \&
  {Conroy}}]{behroozi13}
{Behroozi} P.~S., {Wechsler} R.~H., {Conroy} C., 2013, \apjl, 762, L31

\bibitem[{{Bett} \& {Frenk}(2012)}]{bett12}
{Bett} P.~E., {Frenk} C.~S., 2012, \mnras, 420, 3324

\bibitem[{{Bett} \& {Frenk}(2016)}]{bett16}
{Bett} P.~E., {Frenk} C.~S., 2016, \mnras, 461, 1338

\bibitem[{{Binney}(1977)}]{binney77}
{Binney} J., 1977, \apj, 215, 483

\bibitem[{{Birnboim} \& {Dekel}(2003)}]{bd03}
{Birnboim} Y., {Dekel} A., 2003, \mnras, 345, 349

\bibitem[{{Birnboim}, {Dekel} \& {Neistein}(2007){Birnboim}, {Dekel}, \&
  {Neistein}}]{bdn07}
{Birnboim} Y., {Dekel} A., {Neistein} E., 2007, \mnras, 380, 339

\bibitem[{{Bryan} \& {Norman}(1998)}]{bryan98}
{Bryan} G.~L., {Norman} M.~L., 1998, \apj, 495, 80

\bibitem[{{Butler}, {Obreschkow} \& {Oh}(2017){Butler}, {Obreschkow}, \&
  {Oh}}]{butler17}
{Butler} K.~M., {Obreschkow} D., {Oh} S.-H., 2017, \apjl, 834, L4

\bibitem[{{Cattaneo} \& {Teyssier}(2007)}]{cattaneo07}
{Cattaneo} A., {Teyssier} R., 2007, \mnras, 376, 1547

\bibitem[{{Ceverino}, {Dekel} \& {Bournaud}(2010){Ceverino}, {Dekel}, \&
  {Bournaud}}]{cdb10}
{Ceverino} D., {Dekel} A., {Bournaud} F., 2010, \mnras, 404, 2151

\bibitem[{{Ceverino} {et~al}\mbox{.}(2012){Ceverino}, {Dekel}, {Mandelker},
  {Bournaud}, {Burkert}, {Genzel}, \& {Primack}}]{ceverino12}
{Ceverino} D., {Dekel} A., {Mandelker} N., {Bournaud} F., {Burkert} A.,
  {Genzel} R., {Primack} J., 2012, \mnras,

\bibitem[{{Ceverino} {et~al}\mbox{.}(2015){Ceverino}, {Dekel}, {Tweed}, \&
  {Primack}}]{ceverino15_e}
{Ceverino} D., {Dekel} A., {Tweed} D., {Primack} J., 2015, \mnras, 447, 3291

\bibitem[{{Ceverino} \& {Klypin}(2009)}]{ceverino09}
{Ceverino} D., {Klypin} A., 2009, \apj, 695, 292

\bibitem[{{Ceverino} {et~al}\mbox{.}(2014){Ceverino}, {Klypin}, {Klimek},
  {Trujillo-Gomez}, {Churchill}, {Primack}, \& {Dekel}}]{ceverino14}
{Ceverino} D., {Klypin} A., {Klimek} E.~S., {Trujillo-Gomez} S., {Churchill}
  C.~W., {Primack} J., {Dekel} A., 2014, \mnras, 442, 1545

\bibitem[{{Ceverino}, {Primack} \& {Dekel}(2015){Ceverino}, {Primack}, \&
  {Dekel}}]{ceverino15_shape}
{Ceverino} D., {Primack} J., {Dekel} A., 2015, \mnras, 453, 408

\bibitem[{{Chabrier}(2003)}]{chabrier03}
{Chabrier} G., 2003, \pasp, 115, 763

\bibitem[{{Chowdhury} \& {Chengalur}(2017)}]{chowdhury17}
{Chowdhury} A., {Chengalur} J.~N., 2017, \mnras, 467, 3856

\bibitem[{{Ciotti} \& {Ostriker}(2007)}]{ciotti07}
{Ciotti} L., {Ostriker} J.~P., 2007, \apj, 665, 1038

\bibitem[{{Cortese} {et~al}\mbox{.}(2014){Cortese}, {Fogarty}, {Ho}, {Bekki},
  {Bland-Hawthorn}, {Colless}, {Couch}, {Croom}, {Glazebrook}, {Mould},
  {Scott}, {Sharp}, {Tonini}, {Allen}, {Bloom}, {Bryant}, {Cluver}, {Davies},
  {Drinkwater}, {Goodwin}, {Green}, {Kewley}, {Kostantopoulos}, {Lawrence},
  {Mahajan}, {Medling}, {Owers}, {Richards}, {Sweet}, \& {Wong}}]{cortese14}
{Cortese} L. {et~al.}, 2014, \apjl, 795, L37

\bibitem[{{Croton} {et~al}\mbox{.}(2006){Croton}, {Springel}, {White}, {De
  Lucia}, {Frenk}, {Gao}, {Jenkins}, {Kauffmann}, {Navarro}, \&
  {Yoshida}}]{croton06}
{Croton} D.~J. {et~al.}, 2006, \mnras, 365, 11

\bibitem[{{Danovich} {et~al}\mbox{.}(2015){Danovich}, {Dekel}, {Hahn},
  {Ceverino}, \& {Primack}}]{danovich15}
{Danovich} M., {Dekel} A., {Hahn} O., {Ceverino} D., {Primack} J., 2015,
  \mnras, 449, 2087

\bibitem[{{Dekel} \& {Birnboim}(2006)}]{db06}
{Dekel} A., {Birnboim} Y., 2006, \mnras, 368, 2

\bibitem[{{Dekel} \& {Birnboim}(2008)}]{db08}
{Dekel} A., {Birnboim} Y., 2008, \mnras, 383, 119

\bibitem[{{Dekel} {et~al}\mbox{.}(2009){Dekel}, {Birnboim}, {Engel},
  {Freundlich}, {Goerdt}, {Mumcuoglu}, {Neistein}, {Pichon}, {Teyssier}, \&
  {Zinger}}]{dekel09}
{Dekel} A. {et~al.}, 2009, \nat, 457, 451

\bibitem[{{Dekel} \& {Burkert}(2014)}]{db14}
{Dekel} A., {Burkert} A., 2014, \mnras, 438, 1870

\bibitem[{{Dekel} {et~al}\mbox{.}(2020){Dekel}, {Ginzburg}, {Jiang},
  {Freundlich}, {Lapiner}, \& {et al.}}]{dekel20_ring}
{Dekel} A., {Ginzburg} O., {Jiang} F., {Freundlich} J., {Lapiner} S., {et al.},
  2020, arXiv e-prints

\bibitem[{{Dekel} \& {Krumholz}(2013)}]{dk13}
{Dekel} A., {Krumholz} M.~R., 2013, \mnras, 432, 455

\bibitem[{{Dekel}, {Lapiner} \& {Dubois}(2019){Dekel}, {Lapiner}, \&
  {Dubois}}]{dekel19_gold}
{Dekel} A., {Lapiner} S., {Dubois} Y., 2019, arXiv e-prints

\bibitem[{{Dekel} \& {Mandelker}(2014)}]{dm14}
{Dekel} A., {Mandelker} N., 2014, \mnras, 444, 2071

\bibitem[{{Dekel}, {Sari} \& {Ceverino}(2009){Dekel}, {Sari}, \&
  {Ceverino}}]{dsc09}
{Dekel} A., {Sari} R., {Ceverino} D., 2009, \apj, 703, 785

\bibitem[{{Dekel} {et~al}\mbox{.}(2019){Dekel}, {Sarkar}, {Jiang}, {Bournaud},
  {Krumholz}, {Ceverino}, \& {Primack}}]{dekel19_ks}
{Dekel} A., {Sarkar} K.~C., {Jiang} F., {Bournaud} F., {Krumholz} M.~R.,
  {Ceverino} D., {Primack} J.~R., 2019, \mnras, 488, 4753

\bibitem[{{Dekel} \& {Silk}(1986)}]{ds86}
{Dekel} A., {Silk} J., 1986, \apj, 303, 39

\bibitem[{{Dekel} \& {Woo}(2003)}]{dw03}
{Dekel} A., {Woo} J., 2003, \mnras, 344, 1131

\bibitem[{{Dekel} {et~al}\mbox{.}(2013){Dekel}, {Zolotov}, {Tweed}, {Cacciato},
  {Ceverino}, \& {Primack}}]{dekel13}
{Dekel} A., {Zolotov} A., {Tweed} D., {Cacciato} M., {Ceverino} D., {Primack}
  J.~R., 2013, \mnras, 435, 999

\bibitem[{{Dubois} {et~al}\mbox{.}(2011){Dubois}, {Devriendt}, {Teyssier}, \&
  {Slyz}}]{dubois11}
{Dubois} Y., {Devriendt} J., {Teyssier} R., {Slyz} A., 2011, \mnras, 417, 1853

\bibitem[{{Dubois} {et~al}\mbox{.}(2015){Dubois}, {Volonteri}, {Silk},
  {Devriendt}, {Slyz}, \& {Teyssier}}]{dubois15}
{Dubois} Y., {Volonteri} M., {Silk} J., {Devriendt} J., {Slyz} A., {Teyssier}
  R., 2015, \mnras, 452, 1502

\bibitem[{{Eisenstein} \& {Hu}(1999)}]{eisenstein99}
{Eisenstein} D.~J., {Hu} W., 1999, \apj, 511, 5

\bibitem[{{El-Badry} {et~al}\mbox{.}(2018{\natexlab{a}}){El-Badry}, {Bradford},
  {Quataert}, {Geha}, {Boylan-Kolchin}, {Weisz}, {Wetzel}, {Hopkins}, {Chan},
  {Fitts}, {Kere{\v s}}, \& {Faucher-Gigu{\`e}re}}]{elbadry18b}
{El-Badry} K. {et~al.}, 2018{\natexlab{a}}, \mnras, 477, 1536

\bibitem[{{El-Badry} {et~al}\mbox{.}(2018{\natexlab{b}}){El-Badry}, {Quataert},
  {Wetzel}, {Hopkins}, {Weisz}, {Chan}, {Fitts}, {Boylan-Kolchin}, {Kere{\v
  s}}, {Faucher-Gigu{\`e}re}, \& {Garrison-Kimmel}}]{elbadry18a}
{El-Badry} K. {et~al.}, 2018{\natexlab{b}}, \mnras, 473, 1930

\bibitem[{{Ferland} {et~al}\mbox{.}(1998){Ferland}, {Korista}, {Verner},
  {Ferguson}, {Kingdon}, \& {Verner}}]{ferland98}
{Ferland} G.~J., {Korista} K.~T., {Verner} D.~A., {Ferguson} J.~W., {Kingdon}
  J.~B., {Verner} E.~M., 1998, \pasp, 110, 761

\bibitem[{{F{\"o}rster Schreiber} {et~al}\mbox{.}(2018){F{\"o}rster Schreiber},
  {Renzini}, {Mancini}, {Genzel}, {Bouch{\'e}}, {Cresci}, {Hicks}, \& {et
  al.}}]{forster18a}
{F{\"o}rster Schreiber} N.~M., {Renzini} A., {Mancini} C., {Genzel} R.,
  {Bouch{\'e}} N., {Cresci} G., {Hicks} E.~K.~S., {et al.}, 2018, \apjs, 238,
  21

\bibitem[{{Freundlich} {et~al}\mbox{.}(2019){Freundlich}, {Combes}, {Tacconi},
  {Genzel}, {Garcia-Burillo}, {Neri}, {Contini}, {Bolatto}, {Lilly},
  {Salom{\'e}}, {Bicalho}, {Boissier}, {Boone}, {Bouch{\'e}}, {Bournaud},
  {Burkert}, {Carollo}, {Cooper}, {Cox}, {Feruglio}, {F{\"o}rster Schreiber},
  {Juneau}, {Lippa}, {Lutz}, {Naab}, {Renzini}, {Saintonge}, {Sternberg},
  {Walter}, {Weiner}, {Wei{\ss}}, \& {Wuyts}}]{freundlich19_PHIBSS2}
{Freundlich} J. {et~al.}, 2019, \aap, 622, A105

\bibitem[{{Gammie}(2001)}]{gammie01}
{Gammie} C.~F., 2001, \apj, 553, 174

\bibitem[{{Genzel} {et~al}\mbox{.}(2008){Genzel}, {Burkert}, {Bouch{\'e}},
  {Cresci}, {F{\"o}rster Schreiber}, {Shapley}, {Shapiro}, {Tacconi}, \& {et
  al.,}}]{genzel08}
{Genzel} R. {et~al.}, 2008, \apj, 687, 59

\bibitem[{{Genzel} {et~al}\mbox{.}(2014){Genzel}, {F{\"o}rster Schreiber},
  {Lang}, {Tacchella}, {Tacconi}, {Wuyts}, \& {et al.}}]{genzel14_rings}
{Genzel} R., {F{\"o}rster Schreiber} N.~M., {Lang} P., {Tacchella} S.,
  {Tacconi} L.~J., {Wuyts} S., {et al.}, 2014, \apj, 785, 75

\bibitem[{{Genzel} {et~al}\mbox{.}(2017){Genzel}, {Schreiber}, {{\"U}bler},
  {Lang}, {Naab}, {Bender}, {Tacconi}, {Wisnioski}, {Wuyts}, {Alexander},
  {Beifiori}, {Belli}, {Brammer}, {Burkert}, {Carollo}, {Chan}, {Davies},
  {Fossati}, {Galametz}, {Genel}, {Gerhard}, {Lutz}, {Mendel}, {Momcheva},
  {Nelson}, {Renzini}, {Saglia}, {Sternberg}, {Tacchella}, {Tadaki}, \&
  {Wilman}}]{genzel17}
{Genzel} R. {et~al.}, 2017, \nat, 543, 397

\bibitem[{{Genzel} {et~al}\mbox{.}(2015){Genzel}, {Tacconi}, {Lutz},
  {Saintonge}, {Berta}, {Magnelli}, {Combes}, {Garc{\'{\i}}a-Burillo}, {Neri},
  {Bolatto}, {Contini}, {Lilly}, {Boissier}, {Boone}, {Bouch{\'e}}, {Bournaud},
  {Burkert}, {Carollo}, {Colina}, {Cooper}, {Cox}, {Feruglio}, {F{\"o}rster
  Schreiber}, {Freundlich}, {Gracia-Carpio}, {Juneau}, {Kovac}, {Lippa},
  {Naab}, {Salome}, {Renzini}, {Sternberg}, {Walter}, {Weiner}, {Weiss}, \&
  {Wuyts}}]{genzel15}
{Genzel} R. {et~al.}, 2015, \apj, 800, 20

\bibitem[{{Guo} {et~al}\mbox{.}(2015){Guo}, {Ferguson}, {Bell}, {Koo},
  {Conselice}, {Giavalisco}, {Kassin}, {Lu}, {Lucas}, {Mandelker}, {McIntosh},
  {Primack}, {Ravindranath}, {Barro}, {Ceverino}, {Dekel}, {Faber}, {Fang},
  {Koekemoer}, {Noeske}, {Rafelski}, \& {Straughn}}]{guo15}
{Guo} Y. {et~al.}, 2015, \apj, 800, 39

\bibitem[{{Guo} {et~al}\mbox{.}(2018){Guo}, {Rafelski}, {Bell}, {Conselice},
  {Dekel}, {Faber}, {Giavalisco}, {Koekemoer}, {Koo}, {Lu}, {Mandelker},
  {Primack}, {Ceverino}, {de Mello}, {Ferguson}, {Hathi}, {Kocevski}, {Lucas},
  {P{\'e}rez-Gonz{\'a}lez}, {Ravindranath}, {Soto}, {Straughn}, \&
  {Wang}}]{guo18}
{Guo} Y. {et~al.}, 2018, \apj, 853, 108

\bibitem[{{Haardt} \& {Madau}(1996)}]{haardt96}
{Haardt} F., {Madau} P., 1996, \apj, 461, 20

\bibitem[{{Hopkins}, {Quataert} \& {Murray}(2012){Hopkins}, {Quataert}, \&
  {Murray}}]{hopkins12b}
{Hopkins} P.~F., {Quataert} E., {Murray} N., 2012, \mnras, 421, 3522

\bibitem[{{Huertas-Company} {et~al}\mbox{.}(2018){Huertas-Company}, {Primack},
  {Dekel}, {Koo}, {Lapiner}, {Ceverino}, {Simons}, {Snyder}, {Bernardi},
  {Chen}, {Dom{\'{\i}}nguez-S{\'a}nchez}, {Lee}, {Margalef-Bentabol}, \&
  {Tuccillo}}]{huertas18}
{Huertas-Company} M. {et~al.}, 2018, \apj, 858, 114

\bibitem[{{Jiang} {et~al}\mbox{.}(2019){Jiang}, {Dekel}, {Kneller}, {Lapiner},
  {Ceverino}, {Primack}, {Faber}, {Macci{\`o}}, {Dutton}, {Genel}, \&
  {Somerville}}]{jiang19_spin}
{Jiang} F. {et~al.}, 2019, \mnras, 488, 4801

\bibitem[{{Kassin} {et~al}\mbox{.}(2012){Kassin}, {Weiner}, {Faber}, {Gardner},
  {Willmer}, {Coil}, {Cooper}, {Devriendt}, {Dutton}, {Guhathakurta}, {Koo},
  {Metevier}, {Noeske}, \& {Primack}}]{kassin12}
{Kassin} S.~A. {et~al.}, 2012, \apj, 758, 106

\bibitem[{{Kere{\v s}} {et~al}\mbox{.}(2005){Kere{\v s}}, {Katz}, {Weinberg},
  \& {Dav{\'e}}}]{keres05}
{Kere{\v s}} D., {Katz} N., {Weinberg} D.~H., {Dav{\'e}} R., 2005, \mnras, 363,
  2

\bibitem[{{Khochfar} \& {Ostriker}(2008)}]{khochfar08}
{Khochfar} S., {Ostriker} J.~P., 2008, \apj, 680, 54

\bibitem[{{Komatsu} {et~al}\mbox{.}(2009){Komatsu}, {Dunkley}, {Nolta},
  {Bennett}, {Gold}, {Hinshaw}, {Jarosik}, \& {et al.}}]{komatsu09}
{Komatsu} E., {Dunkley} J., {Nolta} M.~R., {Bennett} C.~L., {Gold} B.,
  {Hinshaw} G., {Jarosik} N., {et al.}, 2009, \apjs, 180, 330

\bibitem[{{Kraljic}, {Dav{\'e}} \& {Pichon}(2020){Kraljic}, {Dav{\'e}}, \&
  {Pichon}}]{kraljic20}
{Kraljic} K., {Dav{\'e}} R., {Pichon} C., 2020, \mnras, 492, 237

\bibitem[{{Kravtsov}(2003)}]{krav03}
{Kravtsov} A.~V., 2003, \apjl, 590, L1

\bibitem[{{Kravtsov}, {Klypin} \& {Khokhlov}(1997){Kravtsov}, {Klypin}, \&
  {Khokhlov}}]{krav97}
{Kravtsov} A.~V., {Klypin} A.~A., {Khokhlov} A.~M., 1997, \apjs, 111, 73

\bibitem[{{Krumholz}, {Dekel} \& {McKee}(2012){Krumholz}, {Dekel}, \&
  {McKee}}]{kdm12}
{Krumholz} M.~R., {Dekel} A., {McKee} C.~F., 2012, \apj, 745, 69

\bibitem[{{Krumholz} \& {Thompson}(2013)}]{krum_thom13}
{Krumholz} M.~R., {Thompson} T.~A., 2013, \mnras, 434, 2329

\bibitem[{{Mandelker} {et~al}\mbox{.}(2017){Mandelker}, {Dekel}, {Ceverino},
  {DeGraf}, {Guo}, \& {Primack}}]{mandelker17}
{Mandelker} N., {Dekel} A., {Ceverino} D., {DeGraf} C., {Guo} Y., {Primack} J.,
  2017, \mnras, 464, 635

\bibitem[{{Mandelker} {et~al}\mbox{.}(2014){Mandelker}, {Dekel}, {Ceverino},
  {Tweed}, {Moody}, \& {Primack}}]{mandelker14}
{Mandelker} N., {Dekel} A., {Ceverino} D., {Tweed} D., {Moody} C.~E., {Primack}
  J., 2014, \mnras, 443, 3675

\bibitem[{{Moody} {et~al}\mbox{.}(2014){Moody}, {Guo}, {Mandelker}, {Ceverino},
  {Mozena}, {Koo}, {Dekel}, \& {Primack}}]{moody14}
{Moody} C.~E., {Guo} Y., {Mandelker} N., {Ceverino} D., {Mozena} M., {Koo}
  D.~C., {Dekel} A., {Primack} J., 2014, \mnras, 444, 1389

\bibitem[{{Moster} {et~al}\mbox{.}(2010){Moster}, {Somerville}, {Maulbetsch},
  {van den Bosch}, {Macci{\`o}}, {Naab}, \& {Oser}}]{moster10}
{Moster} B.~P., {Somerville} R.~S., {Maulbetsch} C., {van den Bosch} F.~C.,
  {Macci{\`o}} A.~V., {Naab} T., {Oser} L., 2010, \apj, 710, 903

\bibitem[{{Murray}, {Quataert} \& {Thompson}(2010){Murray}, {Quataert}, \&
  {Thompson}}]{murray10}
{Murray} N., {Quataert} E., {Thompson} T.~A., 2010, \apj, 709, 191

\bibitem[{{Musso} {et~al}\mbox{.}(2018){Musso}, {Cadiou}, {Pichon}, {Codis},
  {Kraljic}, \& {Dubois}}]{musso18}
{Musso} M., {Cadiou} C., {Pichon} C., {Codis} S., {Kraljic} K., {Dubois} Y.,
  2018, \mnras, 476, 4877

\bibitem[{{Neistein} \& {Dekel}(2008)}]{neistein08_m}
{Neistein} E., {Dekel} A., 2008, \mnras, 388, 1792

\bibitem[{{Nelson} {et~al}\mbox{.}(2016){Nelson}, {Genel}, {Pillepich},
  {Vogelsberger}, {Springel}, \& {Hernquist}}]{nelson16}
{Nelson} D., {Genel} S., {Pillepich} A., {Vogelsberger} M., {Springel} V.,
  {Hernquist} L., 2016, \mnras, 460, 2881

\bibitem[{{Nelson} {et~al}\mbox{.}(2013){Nelson}, {Vogelsberger}, {Genel},
  {Sijacki}, {Kere{\v s}}, {Springel}, \& {Hernquist}}]{nelson13}
{Nelson} D., {Vogelsberger} M., {Genel} S., {Sijacki} D., {Kere{\v s}} D.,
  {Springel} V., {Hernquist} L., 2013, \mnras, 429, 3353

\bibitem[{{Noguchi}(1999)}]{noguchi99}
{Noguchi} M., 1999, \apj, 514, 77

\bibitem[{{Obreschkow} \& {Glazebrook}(2014)}]{obreschkow14}
{Obreschkow} D., {Glazebrook} K., 2014, \apj, 784, 26

\bibitem[{{Ocvirk}, {Pichon} \& {Teyssier}(2008){Ocvirk}, {Pichon}, \&
  {Teyssier}}]{ocvirk08}
{Ocvirk} P., {Pichon} C., {Teyssier} R., 2008, \mnras, 390, 1326

\bibitem[{{Okamoto}, {Gao} \& {Theuns}(2008){Okamoto}, {Gao}, \&
  {Theuns}}]{okamoto08}
{Okamoto} T., {Gao} L., {Theuns} T., 2008, \mnras, 390, 920

\bibitem[{{Press} \& {Schechter}(1974)}]{press74}
{Press} W.~H., {Schechter} P., 1974, \apj, 187, 425

\bibitem[{{Rees} \& {Ostriker}(1977)}]{ro77}
{Rees} M.~J., {Ostriker} J.~P., 1977, \mnras, 179, 541

\bibitem[{{Roca-F{\`a}brega} {et~al}\mbox{.}(2018){Roca-F{\`a}brega}, {Dekel},
  {Faerman}, {Gnat}, {Strawn}, {Ceverino}, {Primack}, {Macci{\`o}}, {Dutton},
  {Prochaska}, \& {Stern}}]{roca19}
{Roca-F{\`a}brega} S. {et~al.}, 2018, arXiv e-prints

\bibitem[{{Rodr{\'{\i}}guez-Puebla}
  {et~al}\mbox{.}(2017){Rodr{\'{\i}}guez-Puebla}, {Primack}, {Avila-Reese}, \&
  {Faber}}]{rodriguez17}
{Rodr{\'{\i}}guez-Puebla} A., {Primack} J.~R., {Avila-Reese} V., {Faber} S.~M.,
  2017, \mnras, 470, 651

\bibitem[{{Rodr{\'{\i}}guez-Puebla}
  {et~al}\mbox{.}(2016){Rodr{\'{\i}}guez-Puebla}, {Primack}, {Behroozi}, \&
  {Faber}}]{rodriguez16}
{Rodr{\'{\i}}guez-Puebla} A., {Primack} J.~R., {Behroozi} P., {Faber} S.~M.,
  2016, \mnras, 455, 2592

\bibitem[{{Salim}, {Fang} \& {Rich}(2012){Salim}, {Fang}, \& {Rich}}]{salim12}
{Salim} S., {Fang} J.~J., {Rich} R., 2012, \apj, 753, 134

\bibitem[{{Shakura} \& {Sunyaev}(1973)}]{shakura73}
{Shakura} N.~I., {Sunyaev} R.~A., 1973, \aap, 24, 337

\bibitem[{{Shaviv} \& {Dekel}(2003)}]{sd03}
{Shaviv} N.~J., {Dekel} A., 2003, astro-ph/0305527

\bibitem[{{Sheth} \& {Tormen}(2002)}]{sheth02}
{Sheth} R.~K., {Tormen} G., 2002, \mnras, 329, 61

\bibitem[{{Silk}(1977)}]{silk77}
{Silk} J., 1977, \apj, 211, 638

\bibitem[{{Simons} {et~al}\mbox{.}(2017){Simons}, {Kassin}, {Weiner}, {Faber},
  {Trump}, {Heckman}, {Koo}, {Pacifici}, {Primack}, {Snyder}, \& {de la
  Vega}}]{simons17}
{Simons} R.~C. {et~al.}, 2017, \apj, 843, 46

\bibitem[{{Snyder} {et~al}\mbox{.}(2015){Snyder}, {Lotz}, {Moody}, {Peth},
  {Freeman}, {Ceverino}, {Primack}, \& {Dekel}}]{snyder15}
{Snyder} G.~F., {Lotz} J., {Moody} C., {Peth} M., {Freeman} P., {Ceverino} D.,
  {Primack} J., {Dekel} A., 2015, \mnras, 451, 4290

\bibitem[{{Sugiyama}(1995)}]{sugiyama95}
{Sugiyama} N., 1995, \apjs, 100, 281

\bibitem[{{Tacchella} {et~al}\mbox{.}(2016{\natexlab{a}}){Tacchella}, {Dekel},
  {Carollo}, {Ceverino}, {DeGraf}, {Lapiner}, {Mandelker}, \&
  {Primack}}]{tacchella16_prof}
{Tacchella} S., {Dekel} A., {Carollo} C.~M., {Ceverino} D., {DeGraf} C.,
  {Lapiner} S., {Mandelker} N., {Primack} J.~R., 2016{\natexlab{a}}, \mnras,
  458, 242

\bibitem[{{Tacchella} {et~al}\mbox{.}(2016{\natexlab{b}}){Tacchella}, {Dekel},
  {Carollo}, {Ceverino}, {DeGraf}, {Lapiner}, {Mandelker}, \& {Primack
  Joel}}]{tacchella16_ms}
{Tacchella} S., {Dekel} A., {Carollo} C.~M., {Ceverino} D., {DeGraf} C.,
  {Lapiner} S., {Mandelker} N., {Primack Joel} R., 2016{\natexlab{b}}, \mnras,
  457, 2790

\bibitem[{{Tacconi} {et~al}\mbox{.}(2018){Tacconi}, {Genzel}, {Saintonge},
  {Combes}, {Garc{\'{\i}}a-Burillo}, {Neri}, {Bolatto}, \& {et
  al.}}]{tacconi18}
{Tacconi} L.~J., {Genzel} R., {Saintonge} A., {Combes} F.,
  {Garc{\'{\i}}a-Burillo} S., {Neri} R., {Bolatto} A., {et al.}, 2018, \apj,
  853, 179

\bibitem[{{Tacconi} {et~al}\mbox{.}(2013){Tacconi}, {Neri}, {Genzel}, {Combes},
  {Bolatto}, {Cooper}, {Wuyts}, \& {et al.}}]{tacconi13}
{Tacconi} L.~J., {Neri} R., {Genzel} R., {Combes} F., {Bolatto} A., {Cooper}
  M.~C., {Wuyts} S., {et al.}, 2013, \apj, 768, 74

\bibitem[{{Tollet} {et~al}\mbox{.}(2019){Tollet}, {Cattaneo}, {Macci{\`o}},
  {Dutton}, \& {Kang}}]{tollet19}
{Tollet} {\'E}., {Cattaneo} A., {Macci{\`o}} A.~V., {Dutton} A.~A., {Kang} X.,
  2019, \mnras, 485, 2511

\bibitem[{{Tomassetti} {et~al}\mbox{.}(2016){Tomassetti}, {Dekel}, {Mandelker},
  {Ceverino}, {Lapiner}, {Faber}, {Kneller}, {Primack}, \&
  {Sai}}]{tomassetti16}
{Tomassetti} M. {et~al.}, 2016, \mnras, 458, 4477

\bibitem[{{van den Bosch}, {Burkert} \& {Swaters}(2001){van den Bosch},
  {Burkert}, \& {Swaters}}]{bosch01}
{van den Bosch} F.~C., {Burkert} A., {Swaters} R.~A., 2001, \mnras, 326, 1205

\bibitem[{{van Dokkum} {et~al}\mbox{.}(2015){van Dokkum}, {Nelson}, {Franx},
  {Oesch}, {Momcheva}, {Brammer}, {F{\"o}rster Schreiber}, {Skelton},
  {Whitaker}, {van der Wel}, {Bezanson}, {Fumagalli}, {Illingworth}, {Kriek},
  {Leja}, \& {Wuyts}}]{dokkum15}
{van Dokkum} P.~G. {et~al.}, 2015, \apj, 813, 23

\bibitem[{{Wheeler} {et~al}\mbox{.}(2019){Wheeler}, {Hopkins}, {Pace},
  {Garrison-Kimmel}, {Boylan-Kolchin}, {Wetzel}, {Bullock}, {Kere{\v{s}}},
  {Faucher-Gigu{\`e}re}, \& {Quataert}}]{wheeler19}
{Wheeler} C. {et~al.}, 2019, \mnras, 490, 4447

\bibitem[{{Wheeler} {et~al}\mbox{.}(2017){Wheeler}, {Pace}, {Bullock},
  {Boylan-Kolchin}, {O{\~n}orbe}, {Elbert}, {Fitts}, {Hopkins}, \& {Kere{\v
  s}}}]{wheeler17}
{Wheeler} C. {et~al.}, 2017, \mnras, 465, 2420

\bibitem[{{Zolotov} {et~al}\mbox{.}(2015){Zolotov}, {Dekel}, {Mandelker},
  {Tweed}, {Inoue}, {DeGraf}, {Ceverino}, {Primack}, {Barro}, \&
  {Faber}}]{zolotov15}
{Zolotov} A. {et~al.}, 2015, \mnras, 450, 2327

\end{thebibliography}

%%%%%%%%%%%%%%%%%%%%%%%
\appendix

%%%%%%%%%
\section{The VELA Cosmological Simulations}
\label{sec:app_vela}

The VELA suite consists of hydro-cosmological simulations of 34 moderately
massive galaxies. Full details are presented in \citet{ceverino14,zolotov15}.
This suite has been used to study central issues in the evolution of galaxies
at high redshifts, including compaction to blue nuggets and the trigger of
quenching
\citep{zolotov15,tacchella16_ms,tacchella16_prof},
evolution of global shape
\citep{ceverino15_shape,tomassetti16},
violent disc instability \citep{mandelker14,mandelker17},
OVI in the CGM \citep{roca19},
and galaxy size and angular momentum \citep{jiang19_spin}.
Additional analysis of the same suite of simulations are discussed in
\citet{moody14,snyder15}.
In this appendix we give an overview of the key aspects of the simulations
and their limitations.

%--------------
\subsection{The Cosmological Simulations}

The VELA simulations make use of the Adaptive Refinement Tree (ART) code
\citep{krav97,krav03,ceverino09}, which accurately follows the
evolution of a gravitating N-body system and the Eulerian gas dynamics using
an adaptive mesh refinement approach. The adaptive mesh refinement maximum
resolution is $17-35\pc$ at all times, which is achieved at densities of
$\sim10^{-4}-10^3\cmc$.
Beside gravity and hydrodynamics, the code incorporates physical process
relevant for galaxy formation such as gas cooling by atomic hydrogen and
helium, metal and molecular hydrogen cooling, photoionization heating by the
UV background with partial self-shielding, star formation, stellar mass loss,
metal enrichment of the ISM and stellar feedback. Supernovae and stellar winds
are implemented by local injection of thermal energy as described in
\citet{ceverino09,cdb10} and \citet{ceverino12}. Radiation-pressure
stellar feedback is implemented at a moderate level, following
\citet{dekel13}, as described in \citet{ceverino14}.

\smallskip
Cooling and heating rates are tabulated for a given gas density, temperature,
metallicity and UV background based on the CLOUDY code \citep{ferland98},
assuming a slab of thickness $1\kpc$. A uniform UV background based on the
redshift-dependent \citet{haardt96} model is assumed, except at gas densities
higher than $0.1\cmc$, where a substantially suppressed UV
background is used
($5.9\times10^6 \erg\, {\rm s}^{-1} \cms\, {\rm Hz}^{-1}$)
in order to mimic the partial self-shielding of dense gas, allowing dense gas
to cool down to temperatures of $\sim300$K. The assumed equation of
state is that of an ideal mono-atomic gas. Artificial fragmentation on the cell
size is prevented by introducing a pressure floor, which ensures that the Jeans
scale is resolved by at least 7 cells \citep[see][]{cdb10}.

\smallskip
Star particles form in timesteps of $5 \Myr$ in cells where the gas density
exceeds a threshold of $1~\cmc$ and the temperatures is below
$10^4$K. Most stars ($>90\%$) end up forming at temperatures well
below $10^3$K, and more than half of the stars form near
$300$K in cells where the gas density is higher than $10~\cmc$.
The code implements a stochastic star-formation where a star particle with a
mass of $42\%$ of the gas mass forms with a probability
$P=(\rhog/10^3\cmc)^{1/2}$ but not higher than $0.2$.
This corresponds to a local SFR that crudely mimics
$\drhos \epsf \rhog/\tff$ with $\epsf \sim 0.02$.
A stellar initial mass function of \citet{chabrier03} is assumed.

\smallskip
Thermal feedback that mimics the energy release from stellar winds and
supernova explosions s incorporated as a constant heating rate over
the $40~\Myr$ following star formation.
A velocity kick of $\sim10\kms$ is applied
to $30~\%$ of the newly formed stellar particles -- this enables SN explosions
in lower density regions where the cooling may not overcome the heating without
implementing an artificial shutdown of cooling \citep{ceverino09}.
The code also incorporates the later effects of Type Ia supernova and
stellar mass loss, and it follows the metal enrichment of the ISM.

\smallskip
Radiation pressure is incorporated through the addition of a non-thermal
pressure term to the total gas pressure in regions where ionizing photons
from massive stars are produced and may be trapped. This ionizing radiation
injects momentum in the cells neighbouring massive star particles younger than
$5\Myr$, and whose column density exceeds
$10^{21}\cms$, isotropically pressurizing the star-forming
regions \citep[see more details in][]{agertz13,ceverino14}.

\begin{table*}
\centering
\begin{tabular}{@{}ccccccccccccc}
\multicolumn{13}{c}{{\bf Properties of the VELA galaxies}} \\
\hline
Galaxy & $\Mv$ & $\Ms$ & $\Mg$ & SFR & $\Re$ & $\Rd$ & $\Hd$ & $V_{\rm rot}$ & $\sigma$ & $e$ & $f$ & $a_{\rm fin}$\\
 & $10^{12}\msun$ & $10^{10}\msun$ & $10^{10}\msun$ & $\msun/\yr$ & kpc & kpc & kpc & km/s & km/s & & &\\

\hline
\hline
01 & 0.16 & 0.20 & 0.14 & 0.93 & 2.64 & 5.15 & 2.57 & 66.0 & 50.3 & 0.72 & 0.97 & 0.50 \\
02 & 0.13 & 0.16 & 0.12 & 1.81 & 1.43 & 6.37 & 3.57 & 71.6 & 39.1 & 0.81 & 0.98 & 0.50 \\
03 & 0.14 & 0.38 & 0.08 & 1.41 & 3.67 & 5.21 & 2.34 & 78.3 & 59.7 & 0.75 & 0.96 & 0.50 \\
04 & 0.12 & 0.08 & 0.08 & 1.73 & 0.45 & 5.71 & 2.79 & 23.3 & 54.3 & 0.96 & 0.88 & 0.50 \\
05 & 0.07 & 0.07 & 0.05 & 1.81 & 0.38 & 5.36 & 1.98 & 60.4 & 32.8 & 0.94 & 0.75 & 0.50 \\
06 & 0.55 & 2.14 & 0.33 & 1.05 & 20.60 & 2.53 & 0.42 & 221.1 & 48.1 & 0.56 & 1.00 & 0.37 \\
07 & 0.90 & 5.75 & 0.79 & 2.85 & 18.13 & 12.59 & 2.06 & 285.5 & 71.4 & 0.85 & 1.00 & 0.54 \\
08 & 0.28 & 0.35 & 0.15 & 0.74 & 5.70 & 4.03 & 1.53 & 91.6 & 48.5 & 0.80 & 0.96 & 0.57 \\
09 & 0.27 & 1.03 & 0.29 & 1.74 & 3.57 & 7.34 & 2.12 & 152.9 & 36.0 & 0.99 & 0.85 & 0.40 \\
10 & 0.13 & 0.60 & 0.13 & 0.46 & 3.20 & 4.51 & 1.19 & 137.1 & 40.9 & 0.50 & 0.99 & 0.56 \\
11 & 0.27 & 0.76 & 0.33 & 2.14 & 8.94 & 8.34 & 5.08 & 121.3 & 71.1 & 0.90 & 0.80 & 0.46 \\
12 & 0.27 & 1.95 & 0.20 & 1.13 & 2.70 & 6.53 & 1.72 & 181.2 & 43.7 & 0.97 & 0.78 & 0.44 \\
13 & 0.31 & 0.57 & 0.35 & 2.48 & 4.48 & 9.74 & 4.75 & 131.7 & 41.3 & 0.97 & 0.88 & 0.40 \\
14 & 0.36 & 1.26 & 0.44 & 0.32 & 23.31 & 1.10 & 0.14 & 213.9 & 82.8 & 0.43 & 0.98 & 0.41 \\
15 & 0.12 & 0.51 & 0.08 & 1.07 & 1.35 & 6.26 & 1.08 & 110.2 & 37.9 & 0.80 & 0.98 & 0.56 \\
*16* & 0.50 & 4.09 & 0.50 & 0.61 & 18.46 & 6.05 & 0.99 & 269.4 & 104.8 & 0.37 &
0.98 & *0.24* \\
*17* & 1.13 & 8.48 & 1.11 & 1.36 & 61.37 & 7.70 & 1.10 & 288.6 & 180.6 & 0.43 &
0.99 & *0.31* \\
*19* & 0.88 & 4.49 & 0.57 & 1.22 & 40.46 & 1.55 & 0.12 & 257.2 & 91.9 & 0.70 &
0.99 & *0.29* \\
20 & 0.53 & 3.59 & 0.35 & 1.72 & 5.55 & 9.57 & 2.75 & 235.0 & 62.0 & 0.78 & 1.00 & 0.44 \\
21 & 0.62 & 4.05 & 0.43 & 1.73 & 7.89 & 9.48 & 1.18 & 261.6 & 42.9 & 0.52 & 1.00 & 0.50 \\
22 & 0.49 & 4.40 & 0.25 & 1.31 & 12.00 & 4.70 & 0.40 & 285.6 & 50.3 & 0.48 & 1.00 & 0.50 \\
23 & 0.15 & 0.76 & 0.13 & 1.16 & 3.06 & 6.28 & 1.54 & 133.0 & 49.2 & 0.78 & 0.99 & 0.50 \\
24 & 0.28 & 0.88 & 0.25 & 1.68 & 3.88 & 7.29 & 1.95 & 131.5 & 42.0 & 0.99 & 0.97 & 0.48 \\
25 & 0.22 & 0.69 & 0.08 & 0.73 & 2.29 & 5.70 & 0.82 & 93.9 & 71.3 & 0.80 & 0.99 & 0.50 \\
26 & 0.36 & 1.58 & 0.26 & 0.74 & 9.36 & 5.42 & 1.30 & 179.6 & 65.0 & 0.74 & 1.00 & 0.50 \\
27 & 0.33 & 0.71 & 0.29 & 1.98 & 6.10 & 9.16 & 4.97 & 122.8 & 60.4 & 0.25 & 0.98 & 0.50 \\
28 & 0.20 & 0.18 & 0.21 & 2.32 & 5.54 & 5.66 & 2.97 & 37.3 & 84.6 & 0.92 & 0.63 & 0.50 \\
29 & 0.52 & 2.29 & 0.32 & 1.89 & 16.83 & 7.46 & 0.97 & 185.6 & 108.4 & 0.96 & 0.91 & 0.50 \\
30 & 0.31 & 1.57 & 0.24 & 1.43 & 2.97 & 9.32 & 1.67 & 192.1 & 37.3 & 0.68 & 1.00 & 0.34 \\
*31* & 0.23 & 0.78 & 0.13 & 0.43 & 15.26 & 4.19 & 0.96 & 195.4 & 48.1 & 0.82 &
0.99 & *0.19* \\
32 & 0.59 & 2.66 & 0.43 & 2.58 & 14.86 & 4.98 & 1.06 & 195.4 & 56.4 & 0.84 & 1.00 & 0.33 \\
33 & 0.83 & 4.81 & 0.44 & 1.23 & 32.68 & 4.59 & 0.88 & 262.7 & 114.2 & 0.49 & 0.95 & 0.39 \\
34 & 0.52 & 1.57 & 0.44 & 1.84 & 14.47 & 5.29 & 1.87 & 156.9 & 70.8 & 0.29 & 1.00 & 0.35 \\
*35* & 0.23 & 0.56 & 0.25 & 0.33 & 22.93 & 1.13 & 0.30 & 204.4 & 40.4 & - & - &
*0.22* \\

\hline
\end{tabular}%\newline
\caption{Relevant global properties of the VELA 3 galaxies.
The quantities are quoted at $z=2$ ($a=0.33$) 
or at the final timestep $a_{\rm fin}$ when it is $<0.33$ 
(marked by stars).
$\Mv$ is the total virial mass.
The following four quantities are measured within $0.1\Rv$:
$\Ms$ is the stellar mass,
$\Mg$ is the gas mass,
SFR is the star formation rate,
and $\Re$ is the half-stellar-mass radius.
The disc outer volume, as defined in \citet{mandelker14}, is given by
$\Rd$ and $\Hd$, the disc radius and half height.
$V_{\rm rot}$ and $\sigma$ are the rotation velocity and the radial velocity 
dispersion of the gas.
$e$ and $f$ are the shape parameters of the gas distribution, 
representing the ``elongation" and ``flattening" as defined in \citet{tomassetti16}.
$a_{\rm fin}$ is the expansion factor at the last output.
}
\label{tab:sample}
\end{table*}

\smallskip
The initial conditions for the simulations are based on dark-matter haloes that
were drawn from dissipationless N-body simulations at lower resolution in
cosmological boxes of $15-60\Mpc$. The $\Lambda$CDM cosmological model was
assumed with the WMAP5 values of the cosmological parameters,
$\omm=0.27$, $\oml=0.73$, $\omb=0.045$, $h=0.7$ and
$\sigma_8=0.82$ \citep{komatsu09}. Each halo was selected to have a
given virial mass at $z = 1$ and no ongoing major merger at $z=1$.
This latter criterion eliminated less than $10~\%$ of the haloes, those
that tend to be in a dense, proto-cluster environment at $z\sim1$.
The virial masses at $z=1$ were chosen to be in the range
$\Mv=2\times10^{11}-2\times10^{12}~M_{\odot}$, about a
median of $4.6\times10^{11}~M_{\odot}$. If left in isolation, the median mass
at $z=0$ was intended to be $\sim10^{12}~M_{\odot}$.

\smallskip % limitations
The VELA cosmological simulations are state-of-the-art in terms
of high-resolution adaptive mesh refinement hydrodynamics and the treatment of
key physical processes at the subgrid level.
In particular, they trace the cosmological streams that feed galaxies at high
redshift, including mergers and smooth flows, and they resolve the violent disc
instability that governs high-$z$ disc evolution and bulge formation
\citep{cdb10,ceverino12,ceverino15_e,mandelker14}.
Like in other simulations, the treatments of star formation and feedback
processes are rather simplified. The code may assume a realistic SFR efficiency
per free fall time on the grid scale
but it does not follow in detail the formation of
molecules and the effect of metallicity on SFR.
The feedback is treated in a crude way, where the resolution does not allow
the capture of the Sedov-Taylor phase of supernova bubbles.
The radiative stellar feedback assumed
no infrared trapping, in the spirit of low trapping advocated by
\citet{dk13} based on \citet{krum_thom13},
which makes the radiative feedback weaker than in other simulations
that assume more significant trapping \citep{murray10,hopkins12b}.
Finally, AGN feedback, and feedback associated with cosmic rays and magnetic
fields, are not yet implemented. 
As shown in \citet{ceverino14}, the star formation rates, gas
fractions and stellar-to-halo mass ratios are crudely in the general ballpark 
of estimates deduced from observations,
\adb{with a tendency to underestimate the gas fractions and overestimate the
stellar-to-halo mass ratios by up to a factor of two due to the relatively
weak feedback. This limitation may affect certain quantitative measurements, 
but it is not expected to hurt the predictive power of qualitative physical
phenomena.}

%-----------------
\subsection{The Galaxy Sample and Measurements}
\label{subsec:sample}

The virial and stellar properties of the galaxies
are listed in Table~\ref{tab:sample}.
The virial mass $\Mv$ is the total mass within a sphere of radius
$\Rv$ that encompasses an overdensity of $\Delta(z)=
[18\pi^2-82\oml(z)-39\oml(z)^2]/\omm(z)$,
where $\oml(z)$ and $\omm(z)$ are the cosmological
parameters at $z$ \citep{bryan98,db06}. The stellar mass $\Ms$
is the instantaneous mass in stars within a radius of $0.2\Rv$, 
accounting for past stellar mass loss.

\smallskip
We start the analysis at the cosmological time corresponding to expansion 
factor $a=0.125$ (redshift $z=7$).
As can be seen in Table~\ref{tab:sample}, most galaxies reach $a=0.50$ ($z=1$).
Each galaxy is analyzed at output times separated by a constant interval in
$a$, $\Delta a=0.01$, corresponding at $z=2$ to $\sim100~\Myr$
(roughly half an orbital time at the disc edge).
The sample consists of totally $\sim 1000$ snapshots in the redshift range
$z=7-0.8$ from 35 galaxies that at $z = 2$ span the stellar mass range
$(0.2-6.4)\times10^{11}\Msun$. The half-mass sizes
$\Re$ are determined from the $\Ms$ that are measured within a
radius of $0.2\Rv$ and they range $\Re\simeq0.4-3.2\kpc$ at $z=2$.

\smallskip
The SFR for a simulated galaxy is obtained by
${\rm SFR}=\langle M_{\star}(t_ {\rm age}<t_{\rm max})/t_{\rm max} 
\rangle_{t_{\rm max}}$, where $\Ms(t_{\rm age}<t_{\rm max})$ is the mass
at birth in stars younger than $t_{\rm max}$.
The average $\langle\cdot\rangle_{t_{\rm max}}$ is obtained by averaging over
all $t_{\rm max}$ in the interval $[40,80]\Myr$ in steps of $0.2\Myr$.
The $t_{\rm max}$ in this range are long enough to ensure good statistics.
The SFR ranges from $\sim 1$ to $33 \Msun\yr^{-1}$
at $z\sim2$.

\smallskip
The instantaneous mass of each star particle is derived from its initial mass
at birth and its age using a fitting formula for the mass loss from the stellar
population represented by the star particle,
according to which 10\%, 20\% and 30\% of the mass is lost after
30 Myr, 260 Myr , and 2 Gyr from birth, respectively.
We consistently use here the instantaneous stellar mass, $\Ms$, and
define the specific SFR by
${\rm sSFR}={\rm SFR}/\Ms$.

\smallskip
The determination of the centre of the galaxy is outlined in detail in
Appendix B of \citet{mandelker14}.
Briefly, starting form the most bound star, the centre is refined iteratively
  by calculating the centre of mass of stellar particles in spheres of
decreasing radii, updating the centre and decreasing the radius at each
iteration.  We begin with an initial radius of 600 pc, and decrease the radius
by a factor of 1.1 at each iteration. The iteration terminates when the radius
reaches 130 pc or when the number of stellar particles in the sphere drops
below 20.

\smallskip
%Disk definition from Mandelker. Values in the table}
The disc plane and dimensions are determined iteratively, as detailed in
\citet{mandelker14}. The disc axis is defined by the angular
momentum of cold gas ($T < 1.5 \times 10^{4}$K), which on average accounts for
$\sim 97\%$ of the total gas mass in the disc.
The radius $\Rd$ is chosen to contain $85\%$ of the cold gas mass in the
galactic mid-plane out to $0.15\Rv$, and the half-height $\Hd$ is defined to
encompass $85\%$ of the cold gas mass in a thick cylinder where both the
radius and half-height equal $\Rd$.

\smallskip % table
Relevant global properties of the VELA 3 galaxies at $z=2$ are listed
in \tab{sample} and explained in the caption. It includes the global
masses and sizes of the different components, the shape and kinematic
properties.

\smallskip % BN
We attempt to identify the major event of compaction to a blue nugget 
for each galaxy. This is the one that leads to a significant central gas
depletion and SFR quenching, and marks the transition from dark-matter to
baryon dominance within $\Re$. Following \citet{zolotov15} and
\citet{tacchella16_prof}, the most physical way to identify the compaction
and blue nugget is by the steep rise of gas density (and SFR) within the inner 
$1\kpc$ to the highest peak, as long as it is followed by a significant, 
long-term decline in central gas mass density (and SFR). 
The onset of compaction can also be identified as the start of the steep rise 
of central gas density prior to the blue-nugget peak.
An alternative identification is using the shoulder of the stellar mass density
within $1\kpc$ where its rise due to starburst associated with the compaction
turns into a plateau of maximum long-term compactness slightly after 
the blue-nugget peak of gas density. This is a more practical way to identify
blue nuggets in observations \citep[e.g.][]{barro17}.

\smallskip % mergers
Major mergers in the history of each galaxy, for the limited purpose they are 
used here, are identified in a simplified way by following sudden increases in
the stellar mass.

%%%%%%%%%%%%%%
\section{Complementary figures}
\label{sec:more_figures}

\begin{figure*} % B1 to fig 3
\centering
\includegraphics[width=0.9\textwidth]{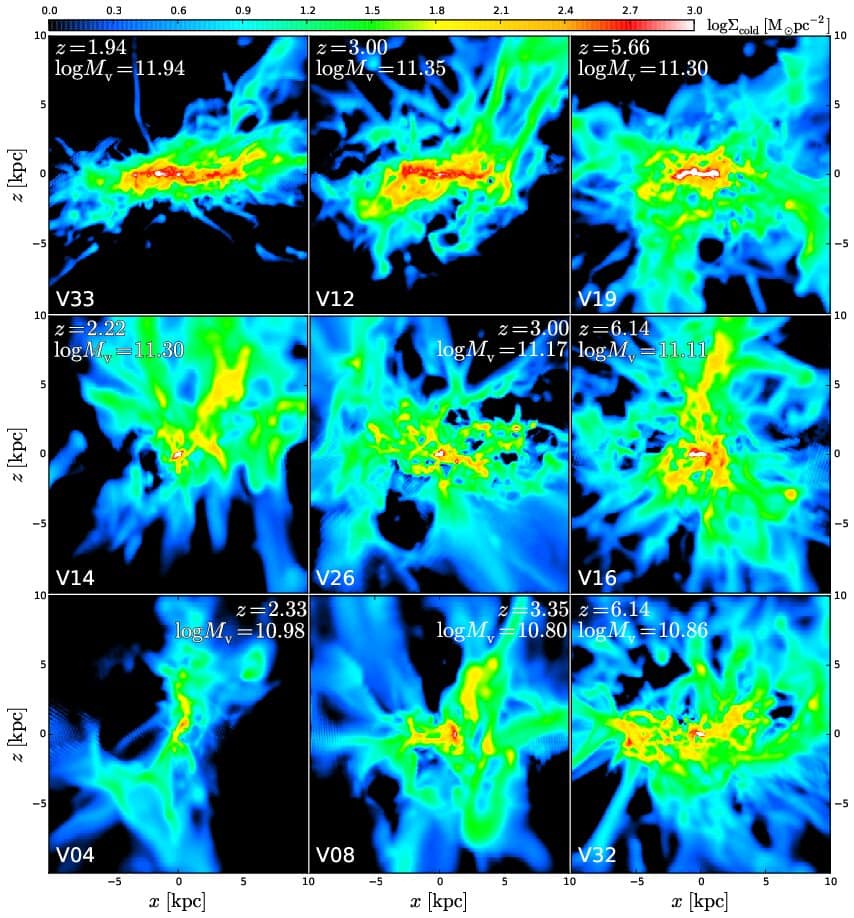}
%\vskip 16cm
%\special{psfile="figs/figB1.ps" %3x3_edgeon_mosaic.ps"
%                 hscale=50 vscale=50   hoffset=40 voffset=0}
\caption{
Projected gas density in example VELA galaxies in different zones in 
the $\Mv-z$ plane, the edge-on counterpart of \fig{Mv-z_face}.
The rows from top to bottom represent galaxies above, near and below
$\Mv=10^{11}\msun$,
while the columns from left to right represent low, intermediate and high
redshifts.
The projections are edge-on with respect to the angular momentum.
}
\label{fig:Mv-z_edge}
\end{figure*}

\begin{figure*} % B2 to fig 3
\centering
\includegraphics[width=0.9\textwidth]{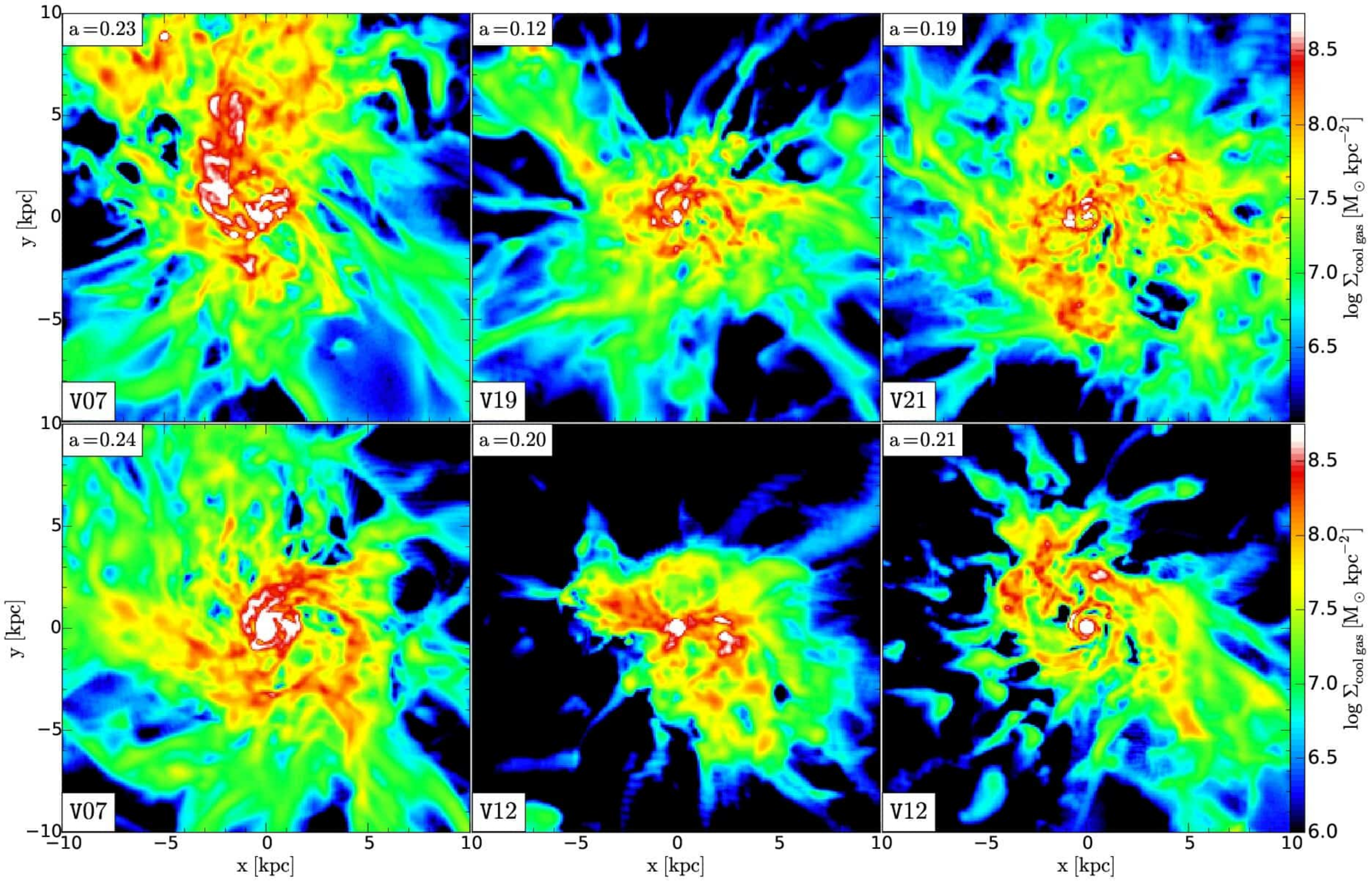}
%\vskip 10.6cm
%\special{psfile="figs/figB2.eps" 
%                          % V_faceon_6panel_below_crit_mass_cool-gas_dpi20.eps"
%                 hscale=25 vscale=25   hoffset=180 voffset=52}
\caption{
Several more examples of pre-compaction non-discs, complementary to
\fig{Mv-z_face}.
Shown are face-on projected images of gas density in six VELA galaxies
at the pre-compaction phase or during the compaction process itself.
They display a typical non-disc perturbed morphology and sometimes showing 
wet mergers in progress.
}
\label{fig:non-discs}
\end{figure*}

Here we show complementary relevant images from the VELA simulations. 

\smallskip
\Fig{Mv-z_edge} is the edge-on analog of \fig{Mv-z_face}.
The galaxies above the critical mass are perturbed, thick, turbulent, rotating
discs, with axial ratios $\sim 4$.
At and below the critical mass there is no evidence for discs, and ongoing
mergers can be seen.

\smallskip
\Fig{non-discs} displays six examples of gas in  galaxies in the 
pre-compaction and compaction phases, showing ongoing wet mergers in progress.

%%%%%%%%%%%%%%%%%%%%%%%%%%%%%%%%%%%%%%%%%%%%%

\label{lastpage}
\end{document}